\documentclass[12pt]{article}
\input psfig.sty

\usepackage{amsmath,amssymb,graphicx} 
\usepackage{cite}

\newcommand{\beq}{\begin{eqnarray}}
\newcommand{\eeq}{\end{eqnarray}}

\newcommand{\centeron}[2]{{\setbox0=\hbox{#1}\setbox1=\hbox{#2}\ifdim
                           \wd1>\wd0\kern.5\wd1\kern-.5\wd0\fi \copy0
                           \kern-.5\wd0\kern-.5\wd1\copy1\ifdim\wd0>\wd1
                           \kern.5\wd0\kern-.5\wd1\fi}}
\newcommand{\ltap}{\>\centeron{\raise.35ex\hbox{$<$}}
                   {\lower.65ex\hbox{$\sim$}}\>}
\newcommand{\gtap}{\>\centeron{\raise.35ex\hbox{$>$}}
                   {\lower.65ex\hbox{$\sim$}}\>}

\newcommand\ZZ{\hbox{\zfont Z\kern-.4emZ}}
\font\zfont = cmss10 
\newcommand{\sfrac}[2]{{\textstyle\frac{#1}{#2}}}

\def\tv#1{\vrule height #1pt depth 5pt width 0pt}

\newcommand{\Circleplus}{{\bf +}}
\newcommand{\Z}{Z}
\newcommand{\lag}{\mathcal{L}}

\newcommand{\arline}{\nonumber \\}

\newcommand{\pd}{\partial}

\begin{document}
\ifx\href\undefined\else\hypersetup{linktocpage=true}\fi 
\begin{titlepage}
\begin{flushright}
{\tt hep-ph/0510275} \\
\end{flushright}

\vspace*{0.8cm}
\begin{center}
\vspace*{0.5cm}
{\LARGE \bf TASI Lectures on Electroweak Symmetry Breaking from
Extra Dimensions\footnote{Lectures at
the Theoretical Advanced Study Institute 2004,
University of Colorado, Boulder, CO June 3-28, 2004.}} \\
\vspace*{1.5cm}

\mbox{\bf
{Csaba Cs\'aki, Jay Hubisz and Patrick Meade}} \\

\vspace*{0.4cm}

{\it Institute of High Energy Phenomenology,\\
Newman Laboratory of Elementary Particle Physics, \\
Cornell University, Ithaca, NY 14853} \\
{\tt  csaki,hubisz,meade@lepp.cornell.edu}
\end{center}

\vspace*{1cm}

\begin{abstract}
\vskip 3pt \noindent This is a pedagogical introduction into the
possible uses and effects of extra dimensions in electroweak (TeV
scale) physics, and in particular to models of electroweak
symmetry breaking via boundary conditions (``higgsless models'').
It is self contained: all the aspects of extra dimensional and
electroweak physics used here are reviewed, before we apply  these
concepts to higgsless models. In the first lecture gauge theories
in an extra dimension and on an interval are discussed. In the
second lecture we describe the basic structure of higgsless
models, while in the third lecture we discuss fermions in extra
dimensions and the inclusion of fermions into higgsless models.
The final lecture is devoted to the issue of electroweak precision
observables in theories beyond the standard model and its
applications to extra dimensional theories and in particular the
higgsless models.
\end{abstract}

\end{titlepage}

\newpage
\tableofcontents


\section{Introduction}
\label{sec:intro}
\setcounter{equation}{0}
\setcounter{footnote}{0}

Theories with extra dimensions have again become very popular over
the past ten years. The reason why is that it was realized that
extra dimensions could actually play an active role in the physics
of the TeV scales (rather than being irrelevant up to the Planck
scale). The first proposal along these lines were theories with
large extra dimensions~\cite{ADD}, where one would explain the
discrepancy of the weak and Planck scale via the presence of large
extra dimensions diluting the strength of gravitational
interactions. The second main wave of excitement was brought on by
the Randall-Sundrum models~\cite{RS1,RS2}, where it was understood
that the geometry of the extra dimensions could actually lead to
novel approaches to the hierarchy problem or even to 4D gravity.
Many of the aspects of these models have been reviewed in a
previous TASI lecture~\cite{tasi02} (and other excellent
introductions to these topics include lectures at this TASI given
by Graham Kribs~\cite{Grahamtasi} and Raman
Sundrum~\cite{Ramantasi}, and can also be found
in~\cite{otherlectures}). The aim of these lectures is to
emphasize less (compared to~\cite{tasi02}) the gravitational
aspects of extra dimensional theories, but rather the use of such
models for electroweak physics. Some knowledge of the material
in~\cite{tasi02} could be useful for reading these notes, however
we have attempted to present a self-contained series of lectures
focusing on gauge theories in extra dimensions.

In the first lecture we discuss generalities about extra
dimensional gauge theories. We mostly focus on the issue of how to
assign a consistent set of boundary conditions (BC's) after proper
gauge fixing. We also comment on the relation between the orbifold
and the interval approaches to describing an extra dimension with
a boundary. In the second lecture we apply the tools from the
first lecture towards building an extra dimensional model where
electroweak symmetry is broken via BC's (rather than by a scalar
higgs). Such models will be referred to as higgsless models, and
will serve throughout these notes as the canonical example of
applying the various concepts discussed. In order to find a (close
to) realistic higgsless model we need to review the basics of the
AdS/CFT correspondence for warped extra dimensions. The third
lecture deals with fermions in extra dimensions. After a brief
review of fermions in general we show how one generically
introduces fermions into extra dimensional models and how theories
with boundaries can render such models chiral. We then show how
these tools can be applied to higgsless models to obtain a
realistic fermion mass spectrum. Finally, we discuss the issue of
how to calculate corrections to electroweak precision observables
in theories beyond the standard model. We give a general effective
field theory approach applicable to any model, and show how the
Peskin-Takeuchi S,T,U formalism fits into it. We then show how
these parameters can be generically calculated in an extra
dimensional model, and evaluate it for the higgsless theory.

\section{Gauge theories in an extra dimension and on an interval}
\label{sec:lec1} \setcounter{equation}{0} \setcounter{footnote}{0}

In this first lecture we will be discussing the structure of a
gauge theory in a single extra dimension.  For now we will be
assuming that there is no non-trivial gravitational background,
that is the extra dimension is flat. However we still need to
discuss what the geometry of the extra dimension is. There are
three distinct possibilities:

\begin{itemize}
\item Totally infinite extra dimension like the other 3+1 dimensions
\item The extra dimension is a half-line, that is infinite in one direction
\item Finite extra dimension: this could either be a circle or an interval
\end{itemize}

For now we will not be dealing with the very interesting
possibility of a half-infinite extra dimension, which could be
phenomenologically viable in the case of localized gravity (the
so-called RS2 model~\cite{RS2}). In most interesting cases one has to deal
with a compactified extra dimension.

The first question we will be discussing in detail is how to
define a field theory with an extra dimension classically.  When
the space is infinite, one usually requires that the fields vanish
when the coordinates go to infinity, that is
\begin{equation}
\psi \to 0 \ {\rm as}\ r\ \to \infty
\end{equation}
However when the space is finite, one does not necessarily need to
impose $\psi \to 0$ as the boundary conditions (BC's). What will
be the possible BC's then? The two possibilities as mentioned
above for the finite extra dimensions are

\begin{itemize}
\item circle ($\equiv$ an interval with the ends identified): in
this case the boundary conditions for the fields are clear, $\Psi
(2\pi R)=\Psi (0)$. \item interval ($\equiv$ the ends are not
identified)
\end{itemize}

In either case, we can just start from the action principle and
see what BC's one can impose that are consistent with the action
principle. As the simplest example let us first discuss the case
of a single scalar field in an extra dimension~\cite{HJMR,CGMPT}.

\subsection{A scalar field on an interval}

We start with a bulk action for the scalar field
\begin{equation}
S_{bulk}=\int d^4x \int_0^{\pi R} \left( \frac{1}{2} \partial^M
\phi
\partial_M \phi -V(\phi )\right)dy,
\end{equation}
where we have assumed that the interval runs between $0$ and $\pi
R$. The coordinates are labelled by $M=0,1,2,3,5$, while Greek
letters $\mu ,\nu ,\ldots$ will denote our usual four dimensions
$0,1,2,3$. We will also assume throughout these lectures that the
signature of the metric is given by
\begin{equation}
g_{MN}=\left( \begin{array}{ccccc} 1 \\ & -1 \\ & & -1 \\ & & & -1 \\ & & & & -1\end{array} \right)
\end{equation}
We will for simplicity first assume that there is no term added on the
boundary of the interval. Let us apply the variational principle to this
theory:
\begin{equation}
\delta S =\int d^4x \int_0^{\pi R} \left(
\partial^M\phi \partial_M \delta \phi -
\frac{\partial V}{\partial \phi} \delta \phi \right) dy.
\end{equation}
Separating out the ordinary 4D coordinates from the fifth
coordinate (and integrating by parts in the ordinary 4D
coordinates, where we apply the usual requirements that the fields
vanish for large distances) we get
\begin{equation}
\delta S = \int d^4x \int_0^{\pi R} dy \left[ -\partial_\mu
\partial^\mu \phi \delta \phi -\frac{\partial V}{\partial \phi}
\delta \phi -\partial_y \phi
\partial_y \delta \phi \right]
\end{equation}
Since we have not yet decided what boundary conditions one wants to impose
we will have to keep the boundary terms when integrating by parts in the
fifth coordinate $y$:
\begin{equation}
\delta S=\int d^4x \int_0^{\pi R} \left[ -\partial_M \partial^M \phi
 -\frac{\partial V}{\partial \phi} \right] \delta \phi
-\left[ \int d^4 x \partial_y \phi \delta \phi \right]_0^{\pi R}.
\end{equation}

To ensure that the variational principle is obeyed, we need
$\delta S=0$, but since this consists of a bulk and a boundary
piece we require:

\begin{itemize}
\item The bulk equation of motion (EOM) $\partial_M \partial^M \phi
= -\frac{\partial V}{\partial \phi}$ as usual
\item The boundary variation needs to also vanish. This implies that one
needs to choose the BC such that
\begin{equation}
\partial_y \phi \delta \phi\vert_{bound} =0.
\label{scalarBC}
\end{equation}
\end{itemize}

We will be calling a boundary condition natural, if it is obtained
by letting the boundary variation of the field $\delta \phi|_{bound}$ to
be arbitrary. In this case the natural BC would be $\partial_y \phi =0$ --
a flat or Neumann BC. But at this stage this is not the only possibility:
one could also satisfy (\ref{scalarBC}) by imposing $\delta \phi|_{bound}=0$
which would follow from the $\phi|_{bound}=0$ Dirichlet BC.
Thus we get two possible BC's for a scalar field on an interval with no
boundary terms:
\begin{itemize}
\item Neumann BC $\partial_y \phi| =0$
\item Dirichlet BC $\phi| =0$
\end{itemize}
However, we would only like to allow the natural boundary
conditions in the theory since these are the ones that will not
lead to explicit (hard) symmetry breaking once more complicated
fields like gauge fields are allowed.  Thus in order to still
allow the Dirichlet BC one needs to reinterpret that as the
natural BC for a theory with additional terms in the Lagrangian
added on the boundary. The simplest possibility is to add  a mass
term to modify the Lagrangian as
\begin{equation}
S=S_{bulk}-\int d^4x \frac{1}{2} M_1^2 \phi^2|_{y=0}-\int d^4x
\frac{1}{2} M_2^2 \phi^2|_{y=\pi R}.
\end{equation}
These will give an additional contribution to the boundary variation
of the action, which will now given by:
\begin{equation}
\delta S_{bound}=-\int \delta \phi (\partial \phi +M_2^2 \phi )|_{y=\pi R}
+\int d^4 x \delta \phi (\partial_y \phi -M_1^2 \phi)|_{y=0}.
\end{equation}
Thus the natural BC's will be given by
\begin{eqnarray}
&& \partial_y \phi +M_2^2 \phi =0 \ \ {\rm at} \ \ y=\pi R, \nonumber \\
&& \partial_y \phi -M_1^2 \phi =0 \ \ {\rm at} \ \ y=0.
\end{eqnarray}
Clearly, for $M_i \to \infty$ we will recover the Dirichlet BC's
in the limit. This is the way we will always understand the Dirichlet BC's:
we will interpret them as the case with infinitely large boundary
induced mass terms for the fields.

Let us now consider what happens in the case~\cite{DGP} when we add a kinetic term on
the boundary (which we will also be calling branes throughout these lectures)
for $\phi$. This is a somewhat tricky question that had many people confused
for a while. For simplicity let us set the mass parameters on the branes
to zero, and take as the action
\begin{equation}
S=S_{bulk}+\int d^4x \frac{1}{2M} \partial_\mu \phi \partial^\mu \phi|_{y=0}.
\end{equation}
Note that the boundary term had to be added with a definite sign, that is
we assume that the arbitrary mass parameter $M$ is positive. This is in
accordance with our expectations that kinetic terms have to have positive signs
if one wants to avoid ghostlike states. For simplicity we have only added a
kinetic term on one of the branes, but of course we could easily repeat the
following analysis for the second brane as well. The boundary variation
at $y=0$ will be modified to
\begin{equation}
\delta S|_0= \int d^4 x \delta \phi (\partial_y \phi-
\frac{1}{M} \Box_4 \phi)|_{y=0}.
\end{equation}
Thus the natural BC will be given by:
\begin{equation}
\partial_y \phi =\frac{1}{M} \Box_4 \phi.
\end{equation}
Using the bulk equation of motion (in the presence of no bulk
potential) $\Box_5 \phi =\Box_4\phi -\phi''=0$ we could also write this
BC as $\phi'=\frac{1}{M} \phi''$. The final form of the BC is obtained by
using the Kaluza-Klein (KK)
decomposition of the field $\phi$ where one usually assumes
that the 4D modes $\phi_n$ have the $x$ dependence $\phi_n(y)
e^{i p_n\cdot x}$, where $p_n^2=m_n^2$ is the n$^{th}$ KK mass eigenvalue.
Using this form the BC will be given by:
\begin{equation}
\partial_y \phi =\frac{1}{M}\Box_4 \phi=-\frac{p_n^2}{M}\phi =
-\frac{m_n^2}{M}\phi.
\end{equation}
In either form this BC is quite peculiar: it depends on the actual mass
eigenvalue in the final form, or involves second derivatives in the first form.
This could be dangerous, since from the theory of differential equations
we know that usually BC's that only involve first derivatives are the
ones that will automatically lead to a hermitian differential operator
on an interval. The usual reason is that the second derivative
operator $d^2/dy^2$ is hermitian if the scalar product
\begin{equation}
(f,g)=\int_0^{\pi R} f^*(y) g(y)
\end{equation}
obeys the relation
\begin{equation}
(f,\frac{d^2}{dy^2} g)= (\frac{d^2}{dy^2}f, g).
\end{equation}
These two terms can be easily transformed into each other using two
integrations by parts up to two boundary terms:
\begin{eqnarray}
&&(f,\frac{d^2}{dy^2} g)=\int_0^{\pi R} f^*(y)\frac{d^2}{dy^2} g(y)=
-\int_0^{\pi R} {f^*}'(y)g'(y)+[f^*(y)g'(y)]_0^{\pi R}= \nonumber \\
&& =\int_0^{\pi R} {f^*}''(y)g(y)+[f^*(y)g'(y)]_0^{\pi
R}-[{f^*}'(y)g(y)]_0^{\pi R}.
\end{eqnarray}
Thus we can see that if the boundary condition for the functions on which
we define this scalar product is of the form
\begin{equation}
f'|_{0,\pi R}=\alpha f|_{0,\pi R}
\end{equation}
then the two boundary terms will cancel each other and the operator
$d^2/dy^2$ is hermitian, and the desired properties (completeness, real
eigenvalues) will automatically follow. However, the boundary condition
$f''=1/M f'$ is not of this form, and the second derivative operator is
naively not hermitian. This is indeed the case, however, one can
choose a different definition of the scalar product on which the above boundary
condition will nevertheless ensure the hermiticity of the second derivative
operator. To find what this scalar product should be, let us try to prove
the orthogonality of two distinct eigenfunctions of the second derivative
operator. Let $f$ and $g$ be two eigenfunctions of the second derivative
operator $f''=\lambda_f f$ and $g''=\lambda_g g$, then
\begin{eqnarray}
&& \int_0^{\pi R} f^* g''dy=\lambda_g \int_0^{\pi R} f^*(y) g(y)
dy= \int_0^{\pi R} f^{*''}(y) g(y) dy +f^{*'}g|_0 -g'f^*|_0.
\end{eqnarray}
Using the boundary condition $f''=M f'=\lambda_f f$, so $f'=\lambda_f /M f$
we find that
\begin{equation}
(\lambda_f-\lambda_g) \left( \int_0^{\pi R} f(y) g(y) dy +\frac{1}{M} fg|_0
\right)=0.
\end{equation}
What we find is the combination that is orthogonal, which means
that this is the combination that one should call the scalar
product in the presence of a non-trivial boundary kinetic term.
Thus the scalar product in this case should be defined by
\begin{equation}
(f,g)=\int_0^{\pi R} f(y) g(y) dy +\frac{1}{M} fg|_0
\end{equation}
This definition still satisfies all the properties that a scalar product should
satisfy, and with this definition the second derivative operator will
now be hermitian. This is easy to check:
\begin{eqnarray}
&& (f,g'')=\int_0^{\pi R} f g'' dy +\frac{1}{M} fg''|_0=-\int_0^{\pi R} f'
g' dy
-fg'|_0+\frac{1}{M} fg|_0=\nonumber \\ && =\int f'' g dy +f'g|_0-f g'|_0 +\frac{1}{M} f g''|_0.
\end{eqnarray}
Due to the BC $g''=M g'$ the last two boundary terms cancel, and the first
can be rewritten as $\frac{1}{M} f'' g|_0$, and so the full
expression really equals $(f'',g)$. So we have shown that there really is
no problem with a theory with brane kinetic terms added. However, one needs to
be careful when using a KK decomposition: the proper scalar product
needs to be used when one is trying to use the orthogonality and
the completeness of the wavefunctions. For example, the completeness
relation will be given by
\begin{equation}
\sum_n g_n(x) g_n(y)=\delta (x-y)-\frac{1}{M}\delta (x) \sum_n g_n(0) g_n(y).
\end{equation}

\subsection{Pure gauge theories on an interval: gauge fixing and BC's}

Next we will consider a pure gauge theory in an extra dimension~\cite{Pilaftsis,CGMPT,Matt}. A
gauge field in 5D $A_M$ contains a 4D gauge field $A_\mu$ and a 4D
scalar $A_5$. The 4D vector will contain a whole KK tower of
massive gauge bosons, however as we will see below the KK tower of
the $A_5$ will be eaten by the massive gauge fields and (except
for a possible zero mode) will be non-physical. That this is what
happens can be guessed from the fact that the Lagrangian contains
a mixing term between the gauge fields and the scalar, reminiscent
of the usual 4D Higgs mechanism. The Lagrangian is given by the
usual form
\begin{equation}
S=\int d^5x (-\frac{1}{4} F_{MN}^a F^{MN\,a} )= \int d^5
x(-\frac{1}{4} F_{\mu \nu}^a F^{\mu\nu\,a}-\frac{1}{2} F_{\mu 5}^a
F^{\mu 5\,a}),
\end{equation}
where the field strength is given by the usual expression
$F_{MN}^a= \partial_M A_N^a-\partial_N A_M^a +g_5 f^{abc} A_M^b A_N^c$.
$g_5$ is the 5D gauge coupling, which has mass dimension $-1/2$, thus
the theory is non-renormalizable, so it has to be considered as a low-energy
effective theory valid below a cutoff scale, that we will be calculating
later on.

To determine the gauge fixing term, let us consider the mixing term between
the scalar and the gauge fields:
\begin{eqnarray}
&& \int d^4x \int_0^{\pi R}dy -\frac{1}{2} F_{\mu 5}^a F^{\mu 5 \
a}|_{quadratic} =\nonumber\\&& \int d^4x \int_0^{\pi R}dy
-\frac{1}{2} (\partial_\mu A_5^a-\partial_5 A_\mu^a) (\partial^\mu
A^{5 \ a}-\partial^5 A^{\mu \ a})= \nonumber \\ && \int d^4x
\int_0^{\pi R}dy -\frac{1}{2} (\partial_\mu A_5^a \partial^\mu
A^{5 \ a}+\partial_5 A_\mu^a
\partial^5 A^{\mu \ a} -2\partial_5 A_\mu^a \partial^\mu A^{5 a}).
\end{eqnarray}
Thus the mixing term that needs to be cancelled is given by
\begin{equation}
\int_0^{\pi R} \partial_5 A_\mu^a \partial^\mu A^{5 a}.
\end{equation}
Integrating by parts we find
\begin{equation}
-\int_0^{\pi R} \partial^\mu A_\mu^a \partial_5 A_5 ^a+[\partial_\mu A^{\mu a}
A_5^a ]_0^{\pi R}.
\label{bulkmixing}
\end{equation}
The bulk term can be cancelled by adding a gauge fixing term of the form
\begin{equation}
S_{GF}=\int d^4x \int_0^{\pi R} -\frac{1}{2\xi }(\partial_\mu
A^{\mu a}-\xi
\partial_5 A_5^a)^2.
\end{equation}
This term is chosen such that the $A_5$ independent piece agrees
with the usual Lorentz gauge fixing term, and such that the cross
term exactly cancels the mixing term from (\ref{bulkmixing}). Thus
within $R_\xi$ gauge, which is what we have defined, the
propagator for the 4D gauge fields will be the usual ones. Varying
the full action we then obtain the bulk equations of motion and
the possible BC's. After integrating by parts we find:
\begin{eqnarray}
&& \delta S_{bulk}=\int d^4x \int_0^{\pi R} \left[ ( \partial_M F^{M\nu a}
-g_5 f^{abc}
F^{M\nu b} A_M^c+\frac{1}{\xi} \partial^\nu\partial^\sigma A_\sigma^a -
\partial^\nu\partial_5 A_5^a)\delta A_\nu^a +\right. \nonumber \\
&& \left.
(\partial^\sigma F^a_{\sigma 5}
-g_5f^{abc}F^b_{\sigma 5}A^{c\sigma}\partial_5\partial_\sigma A^{a\sigma}
-\xi \partial_5^2 A_5^a) \delta A_5^a \right]
\end{eqnarray}
The bulk equations of motion will be that the coefficients of
$\delta A_\nu^a$ and $\delta A_5^a$ in the above equation vanish.
We can see, that the $A_5^a$ field has a term $\xi
\partial_5^2 A_5^a$ in its equation. This will imply that if the
wave function is not flat (e.g. the KK mode is not massless), then
the field is not physical (since in the unitary gauge $\xi \to
\infty$ this field will have an infinite effective 4D mass and
decouples). This shows that as mentioned above, the scalar KK
tower of $A_5^a$ will be completely unphysical due to the 5D Higgs
mechanism, except perhaps for a zero mode for $A_5^a$. Whether or
not there is a zero mode depends on the BC for the $A_5$ field. We
will see later how to interpret $A_5$ zero modes.

In order to eliminate the boundary mixing term in (\ref{bulkmixing}),
we also need to add a boundary gauge fixing term with an a priori unrelated
boundary gauge fixing coefficient $\xi_{bound}$:
\begin{equation}
-\frac{1}{2\xi_{bound}} \int d^4x (\partial_\mu A^{\mu a}\pm
\xi_{bound} A_5^a)^2 |_{0, \pi R},
\end{equation}
where the $-$ sign is for $y=0$ and the $+$ for $y=\pi R$. The boundary
variations are then given by:
\begin{eqnarray}
&& (\partial_5 A^{\mu a} +\frac{1}{\xi_{bound}} \partial_\nu
\partial^\mu A^{\nu a} ) \delta A_\mu^a|_{0,\pi R} +(\xi
\partial_5 A_5^a\pm \xi_{bound} A_5^a)\delta A_5^a|_{0,\pi R}.
\end{eqnarray}
The natural boundary conditions in an arbitrary gauge $\xi, \xi_{bound}$ are
given by
\begin{equation}
\partial_5 A^{\mu a} +\frac{1}{\xi_{bound}} \partial_\mu \partial^\mu
A^{\mu a}=0, \ \ \xi \partial_5 A_5^a\pm \xi_{bound} A_5^a=0.
\end{equation}
This simplifies quite a bit if we go to the unitary gauge on the boundary
given by $\xi_{bound}\to \infty$. In this case we are left with the simple
set of boundary conditions
\begin{equation}
\partial_5 A^{\mu a} =0, \ \ A_5^a=0.
\end{equation}
This is the boundary condition that one usually imposed for gauge
fields in the absence of any boundary terms. Note, that again we
could have chosen some non-natural boundary conditions, where
instead of requiring that the boundary variation be arbitrary we
would require the boundary variation itself (and thus some of the
fields on the boundary) to be vanishing. It turns out that these
boundary conditions would lead to a hard (explicit) breaking of
gauge invariance, and thus we will not consider them in the
following discussion any further. We will see below how these
simple BC's will be modified if one adds scalar fields on the
branes.

\subsection{Gauge theories with boundary scalars}
Let us now consider the case when scalar fields that develop VEV's are added
on the boundary~\cite{CGMPT,Matt}. For simplicity we will be considering a U(1) theory,
but it can be straightforwardly generalized to more complicated groups.
The localized Lagrangians for the two complex scalar
fields $\Phi_i$ will be the usual ones for a Higgs field
in 4D, and the subscripts $i=1,2$ correspond to the two boundaries:
\begin{equation}
{\cal L}_{i}=|D_\mu \Phi_i|^2 -\lambda_i (|\Phi_i|^2-\frac{1}{2}v_i^2)^2.
\end{equation}
These boundary terms will induce non-vanishing VEV's and we
parameterize the Higgs as usual as a physical Higgs and a
Goldstone (pion):
\begin{equation}
\Phi_i=\frac{1}{\sqrt{2}} (v_i+h_i) e^{i \pi_i/v_i}.
\end{equation}
We can now expand again the action to quadratic order in the
fields to find the expression
\begin{eqnarray}
{\cal L}_{4D}=&& \int_0^{\pi R} dy \left( -\frac{1}{4}
F_{\mu\nu}^2 +\frac{1}{2}
(\partial_\mu A_5)^2-\partial_\mu A_5 \partial_5 A^\mu \right) \nonumber \\
&&+\left[\frac{1}{2} (\partial_\mu h_1)^2-\frac{1}{2} \lambda_1 v_1^2 h_1^2
+\frac{1}{2} (\partial_\mu \pi_1-v_1 A_\mu)^2\right]_{y=0}\nonumber \\
&&+\left[\frac{1}{2} (\partial_\mu h_2)^2-\frac{1}{2} \lambda_2 v_2^2 h_2^2
+\frac{1}{2} (\partial_\mu \pi_2-v_2 A_\mu)^2\right]_{y=\pi R}
\end{eqnarray}
Repeating the procedure in the previous section and integrating by
parts, we find that the following bulk and boundary gauge fixing
terms are necessary in order to eliminate all the mixing terms
between the gauge field and the scalars $A_5,\pi_i$:
\begin{eqnarray}
{\cal L}_{GF}=&& -\frac{1}{2\xi} \int_0^{\pi R} dy  (\partial_\mu A^\mu -
\partial_5 A_5)^2-\left[\frac{1}{2\xi_1} (\partial_\mu A^\mu+\xi_1 (v_1 \pi_1
-A_5))^2 \right]_{y=0} \nonumber \\
&& -\left[\frac{1}{2\xi_2} (\partial_\mu A^\mu+\xi_2 (v_2 \pi_2
+A_5))^2 \right]_{y=\pi R}
\end{eqnarray}
With this gauge fixing the gauge field will be decoupled from the rest of
the fields, and its bulk action will be given by
\begin{equation}
\int d^5 x \frac{1}{2} A_\mu \left[ (\partial^2-\partial_y^2)\eta^{\mu \nu}
-(1-\frac{1}{\xi})\partial^\mu \partial^\nu \right] A_\nu.
\end{equation}
The KK decomposition can then be obtained by writing the field as
\begin{equation}
A_\mu (x,y)= \epsilon_\mu (p) e^{i p\cdot x} f(z),
\end{equation}
where $\epsilon_\mu (p)$ is the polarization vector and $f(z)$ is
the wave function. For a given mode we assume that $p^2=m_n^2$.
The equation of motion satisfied by the wave function will simply
be
\begin{equation}
(\partial_y^2+m_n^2) f_n(y)=0,
\end{equation}
which will be linear combinations of sine and cosine functions.
The natural boundary condition (analogous to the previous case)
will now be modified to
\begin{equation}
\partial_y A_\mu \mp v_{1,2}^2 A_\mu =0.
\end{equation}
Thus we can see, that introducing a boundary scalar field will
modify the BC (just like for the case of the simple bulk scalar).
In the limit when $v_i\to \infty$ we simply obtain a Dirichlet BC
for the gauge field. In this limit of a Dirichlet boundary
condition, the fields $h_i,\pi_i$ will clearly decouple from the
gauge field, since they are non-vanishing only where the gauge
field itself vanishes. Thus their effect will be to repel the wave
function of the gauge field from the brane, and to make the gauge
field massive. However, as we will see later even in the limit
$v_i \to \infty$ the mass of the gauge field will not diverge, but
rather it will be given by the radius of the extra dimension.

Finally, let us consider what will happen to the scalar fields and
their BC's. The physical Higgs $h_i$ does not have any mixing term
with any of the other fields, so it will have its own equation of
motion on the branes. Since its mass is determined by the
parameter $\lambda_i$ which does not appear anywhere else in the
theory, we could just make this scalar arbitrarily heavy and
decouple it without influencing any of the other fields. Turning
to the fields $A_5$ and  $\pi_i$ the bulk equation of motion of
$A_5$ will still be given by
\begin{equation}
\partial_y^2 A_5 +\frac{m^2}{\xi} A_5=0.
\end{equation}
Here $m^2$ is the mass of a scalar state that could live in a combination of
$A_5$ and the $\pi_i$'s. The boundary equation of motions for the
$\pi_i$'s will give a relation between these fields and the boundary
values of $A_5$:
\begin{eqnarray}
&& (\frac{m^2}{\xi_1}-v_1^2)\pi_1+v_1 A_5|_{y=0}=0, \nonumber \\
&& (\frac{m^2}{\xi_2}-v_2^2)\pi_2+v_2 A_5|_{y=\pi R}=0.
\label{pis}
\end{eqnarray}
Finally, requiring that the boundary variation for arbitrary field
variations still vanishes (combined with the above two equations)
will give the BC's for the field $A_5$:
\begin{eqnarray}
&& \partial_y A_5 -\frac{\xi_1}{\xi} \frac{m^2/\xi_1}{m^2/\xi_1-v_1^2}
A_5 |_{y=0}=0, \nonumber \\
&& \partial_y A_5 +\frac{\xi_2}{\xi}
\frac{m^2/\xi_2}{m^2/\xi_2-v_2^2} A_5 |_{y=\pi R}=0.
\end{eqnarray}
We can see that when one of the VEV's $v_i$ is turned on, in the
unitary gauge $\xi \to \infty$ the boundary condition for the
$A_5$ field will change from Dirichlet to Neumann BC: $\partial_y
A_5=0$. In the unitary gauge it is also clear that all the massive
modes are again non-physical, since they will provide the
longitudinal components for the massive KK tower of the gauge
fields. However, now it may be possible for physical zero modes in
the scalar fields to exist. Without boundary scalars, the BC for
$A_5$ is Dirichlet, and no non-trivial zero mode may exist. This
basically means that there are just enough many modes in $A_5$ to
provide a longitudinal mode for every massive KK state in the
gauge sector, but no more. When one adds additional scalars on the
boundary, some combination of the $A_5$'s and $\pi$'s may remain
uneaten. If we turn on the VEV for a scalar on both ends (in the
non-ablian case for the same direction), then the $A_5$ will obey
a Neumann BC on both ends and there will be a physical zero mode.
As we will see in the AdS/CFT interpretation this will correspond
to a physical pseudo-Goldstone boson in the theory. In this case
the wave function of the $A_5$ is simply flat (which obeys both
the bulk eom's and the BC's), and from (\ref{pis}) we find that
the boundary scalars will be given by $\pi_i = A_5/v_i$. In the
limit $v_i\to \infty$ the boundary scalars will vanish as expected
and decouple from the theory. If a direction is higgsed only on
one of the boundaries (that is $v_1\neq 0$ but $v_2 =0$) then
there will be no physical scalars in the spectrum. This is the
situation we will be using most of the time in these lectures,
thus in most cases we can simply set all scalar fields to zero and
safely decouple them from the gauge fields in the $v\to \infty$
limit.

\subsection{Orbifold or interval?}

The more traditional way of introducing BC's in theories with
extra dimensions is via the procedure~\cite{stringorbifold} known as ``orbifolding''. By
orbifolding we mean a set of identifications of a geometric
manifold which will reduce the fundamental domain of the theory.

In the case of a single extra dimension, the most general
orbifolding can be described as follows.  Let us first start with
an infinite extra dimension, an infinite line $\mathbb{R}$,
parametrized by $y$, $-\infty <y<\infty$. One can obtain a circle
$S^1$ from the line by the identification $y\to y+2\pi R$, which
we is usually referred to as modding out the infinite line by the
translation $\tau$, $\mathbb{R}\to S^1 = \mathbb{R}/\tau$. This
way we obtain the circle.

Another discrete symmetry that we could use to mod out the line is
a $Z_2$ reflection which takes $y\to -y$. Clearly, under this
reflection the line is mapped to the half-line $R^1\to R^1/Z_2$.
If we apply both discrete projections at the same time, we get the
orbifold $S^1/Z_2$. This orbifold is nothing else but the line
segment between $0$ and $\pi R$.

Let us now see how the fields $\varphi (y)$ that are defined on
the original infinite line $\mathbb{R}$ will behave under these
projections, that is what kind of BC's they will obey. The fields
at the identified points have to be equal, except if there is a
(global or local) symmetry of the Lagrangian. In that case, the
fields at the identified points don't have to be {\it exactly}
equal, but merely equal {\it up to a symmetry} transformation,
since in that case the fields are still physically equal. Thus,
under translations and reflection the fields behave as
\begin{eqnarray}
&& \tau (2\pi R)\varphi (y) =T^{-1} \varphi (y+2\pi R), \\
&& {\cal Z} \varphi (y) = Z \varphi (-y),
\end{eqnarray}
where $T$ and $Z$ are matrices in the field space corresponding to some
symmetry transformation of the action.
This means that we have made the field identifications
\begin{eqnarray}
&& \varphi (y+2\pi R)=T \varphi (y), \\
&& \varphi (-y)=Z \varphi(y).
\end{eqnarray}
Again, $Z$ and $T$ have to be symmetries of the action. However,
$Z$ and $T$ are not completely arbitrary, but they have to satisfy
a consistency condition. We can easily find what this consistency
condition is by considering an arbitrary point at location $y$
within the fundamental domain $0$ and $2 \pi R$,  apply first a
reflection around $0$, ${\cal Z}(0)$, and then a translation by
$2\pi R$, which will take $y$ to $2\pi R -y$. However, there is
another way of connecting these two points using the translations
and the reflections: we can first translate $y$ backwards by $2\pi
R$, which takes $y\to y-2\pi R$, and then reflect around $y=0$,
which will also take the point into $2\pi R -y$. This means that
the translation and reflection satisfy the relation:
\begin{equation}
\tau (2\pi R){\cal Z} (0)={\cal Z} (0) \tau^{-1} (2\pi R).
\end{equation}
When implemented on the fields $\varphi$ this means that we need to have the
relation
\begin{equation}
TZ=ZT^{-1}, {\rm or} \ \ ZTZ=T^{-1}
\label{consistency}
\end{equation}
which is the consistency condition that the field transformations $Z$ and $T$
have to satisfy.

As we have seen, the reflection ${\cal Z}$ is a $Z_2$ symmetry, and so
$Z^2=1$. $T$ is not a $Z_2$ transformation, so $T^2\neq 1$.
However, for non-trivial $T$'s  $T\neq 1$
(T is sometimes called the Scherk-Schwarz-twist) one can always
introduce a combination of $T$ and $Z$ which together act like another
$Z_2$ reflection. We can take the combined transformation
$\tau (2\pi R) {\cal Z}(0)$. This combined transformation takes any point
$y$ into $2\pi R-y$. That means, that it is actually a reflection around
$\pi R$, since if $y=\pi R-x$, then the combined transformation takes it to
$\pi R+x$, so $x\to -x$. So this is a $Z_2$ reflection. And using the
consistency condition (\ref{consistency}) we see that for the
combined field transformation $Z'=TZ$
\begin{equation}
{Z'}^2=(TZ)^2= (TZ)(ZT^{-1})=1,
\end{equation}
so indeed the action of the transformation on the fields is also
acting as another $Z_2$ symmetry.  Thus we have seen that the
description of a generic $S^1/Z_2$ orbifold with non-trivial
Scherk-Schwarz twists can be given as two non-trivial $Z_2$
reflections $Z$ and $Z'$, one which acts around $y=0$ and the
other around $y=\pi R$. These two reflections do not necessarily
commute with each other. A simple geometric picture to visualize
the two reflections is to extend the domain to a circle of
circumference $4\pi R$, with the two reflections acting around
$y=0,2\pi R$ for $Z$ and $\pi R, 3\pi R$ for $Z'$. One can either
use this picture with the fields living over the full circle, or
just living on the fundamental domain between $y=0$ and $2\pi R$.
The two pictures are equivalent.

\begin{figure}
\centerline{\includegraphics[width=0.5\hsize]{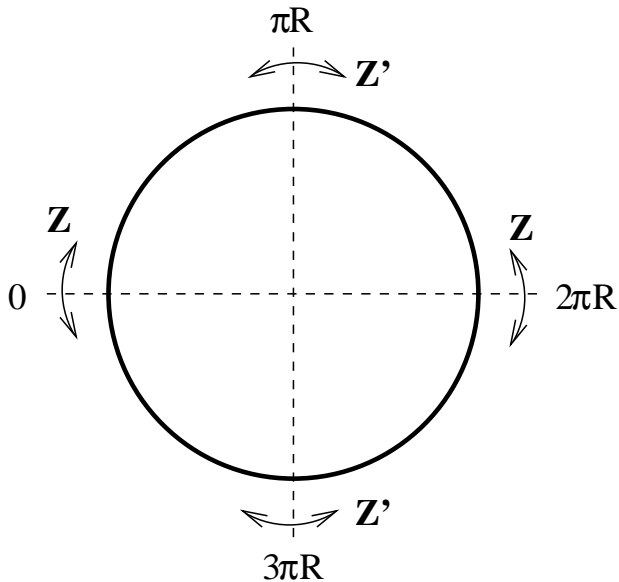}}
\caption{The action of the two $Z_2$ reflections in the extended circle picture. The fundamental domain of the
$S^1/Z_2$ orbifold is just the interval between $0$ and $\pi R$, and the theory can be equivalently
formulated on this line segment as well.}
\end{figure}

So what we find is that in the case of an $S^1/Z_2$ orbifold a field $\varphi (y)$ will live on the
fundamental domain $0<y<\pi R$, and will either have positive or negative parities under the two
$Z_2$ symmetries, which means that it will either have Dirichlet or Neumann BC's at the two boundaries
\begin{equation}
\varphi|_{0,\pi R}=0 \ {\rm or} \ \partial_y \varphi|_{0,\pi R}=0.
\end{equation}
Let us assume that the $Z_2$'s are a subgroup of the original
symmetries of the theory\footnote{The other possibility would be
to use a discrete symmetry of the generators that can not be
expressed as the action of another generator known as an outer
automorphism.  There are few examples of such orbifolds which can
indeed reduce the rank but will not be considered in these
lectures.}, thus in the case of gauge symmetry breaking they need
to be a subgroup of the gauge symmetries. Since it is a $Z_2$
subgroup it is necessarily abelian, so it is a subgroup of the
Cartan subalgebra. This means that the symmetry with which we are
orbifolding commutes at least with the Cartan subalgebra (since it
is a subgroup of the Cartan subalgebra itself), so it will never
reduce the rank of the gauge group. This would imply that the
interval approach would give more general BC's than the orbifolds.
However, one can of course in addition consider orbifolds with
fields localized at the fixed points. With these you can also
reproduce the more general BC's that arise as natural ones from
the interval approach. However, these are more awkward to deal
with since one has to often deal with fields that have
discontinuities (jumps) at the fixed points. So while the two
approaches are nominally equivalent, the interval approach is by
far easier to deal with when the BC's are complicated. The
interval approach is also more advantageous since it allows for a
dynamical explanation of the BC's. One usual
application~\cite{HN,KawaAF} of orbifold theories for example is
to break an SU(5) GUT symmetry to the SM subgroup SU(3)$\times
$SU(2)$\times $U(1), without a Higgs field, and thus
avoiding the doublet-triplet splitting problem. The
doublet-triplet splitting problem is the question of why in an
SU(5) GUT theory (usually a supersymmetric one) the Higgs doublets
are light (of the order of the weak scale) while triplet that is
necessary to make it a complete SU(5) multiplet has to be heavy
(of the order of the GUT scale) in order to obtain unification of
couplings (and to avoid proton decay in SUSY models). In an
orbifold theory one can simply assume that the gauge fields
(which form an adjoint {\bf 24} of SU(5) ) have the following
$Z_2$ parities at one of the boundary:
\begin{equation}
A_\mu^a \to \left( \begin{array}{ccc|c}
&&& \\ & + & & - \\
&&& \\ \hline & - & & + \end{array}
\right), \ \ A_5^a \to \left( \begin{array}{ccc|c}
&&& \\ & - & & + \\
&&& \\ \hline & + & & - \end{array}
\right).
\label{SU5BC}
\end{equation}
If we take the point of view that the orbifold is the fundamental
object, then one never has to mention the doublet triplet splitting problem.
However a different starting point could be an SU(5) theory on an interval,
with some boundary conditions (that are caused by some dynamics of the
fields on the boundary). To obtain the BC's used in the orbifold
picture one could for example consider the SU(5) theory with
an adjoint scalar on the boundary,
which has a VEV
\begin{equation}
\langle \Sigma \rangle =v \left( \begin{array}{ccccc} 1 \\ & 1\\ & & 1\\ & & & -\frac{3}{2} \\
& & & & -\frac{3}{2} \end{array} \right),
\end{equation}
and then the limit $v\to \infty$ will give the orbifold BC's from
(\ref{SU5BC}) for the gauge fields. In order to also solve the doublet triplet
splitting problem in this picture,
one needs to require that the Higgs (which is a $5$ of SU(5)) has $Z_2$
parity at the same boundary:
\begin{equation}
H={\bf 5} =\left( \begin{array}{c} \\ - \\ \\ \hline + \end{array} \right).
\end{equation}
If this is also coming from some dynamics in the interval picture,
then one would have to assume the presence of a boundary
Lagrangian
\begin{equation}
H^* \Sigma H +m H^* H,
\end{equation}
where in order to obtain the above parities one needs to assume
that $-3/2 v+m=0$. This condition is equivalent to the usual fine
tuning solution in supersymmetric GUTs. So from the interval point of view
the doublet-triplet splitting problem would mainifest itself in the
following way: the 3-2-1 invariant boundary conditions for the Higgs
field are
\begin{equation}
\partial_y H_3+m_3 H_3 =0, \ \ \partial_y H_2+m_2 H_2 =0.
\end{equation}
In order to obtain doublet-triplet splitting we need $m_2/m_3 \sim
10^{-14}$, and thus the set of BCs solving the doublet-triplet
splitting problem is equivalent to the usual fine-tuning problem
of SUSY GUTs. Thus in this picture the doublet-triplet splitting problem
would not really be resolved, but rather just hidden behind the question of
what dynamics will cause these fields to have the above BC's. Whether
or not the doublet-triplet splitting problem is really resolved then
depends on how the extra dimension is really emerging: if some string
theory compactification naturally yields the $S^1/Z_2$ orbifold as its vacuum
state with the necessary parities then the doublet-triplet splitting
problem would be indeed resolved. If however the BC's are due to some
boundary dynamics as discussed above the problem would reemerge. Thus
one can not really decide which interpretation is the right one purely
from the low-energy effective theory, but some knowledge of the UV theory
would be necessary.

\section{Higgsless models of electroweak symmetry breaking}
\label{sec:lec2} \setcounter{equation}{0} \setcounter{footnote}{0}

We have shown above how to find the BC's for a general gauge
theory in an extra dimension. We would like to use now this
knowledge to construct a model of electroweak symmetry breaking,
where the electroweak symmetry is broken by boundary conditions,
and without the presence of a physical scalar Higgs boson in the
theory. First we want to show how the presence of extra dimensions
can postpone the unitarity violation scale in a theory with
massive gauge bosons but without a Higgs scalar~\cite{CGMPT,xdunitarity}.
Then we will show
how to find the simplest model with a massive W and Z bosons
without a scalar Higgs from an extra dimensional model with a flat
extra dimension. We will see that in this model the prediction for
the ratio of the W/Z mass is far from the SM value, which is due
to the absence of a custodial SU(2) symmetry protecting this
ratio~\cite{ADMS}. We will show that such a global symmetry indeed predicts
the right W/Z mass ratio and then following the suggestion of~\cite{ADMS}
use the AdS/CFT correspondence
to build an extra dimensional model incorporating custodial
SU(2)~\cite{CGPT,Nomura}. For other aspects of higgsless models
see~\cite{BPR,CGHST,DHLR1,BN,CCGT,DHLR2,BPRS,MSU,corrections,Howard,Howard2,Maxim,BMP,fermioncollider,Papucci,unitarityfermion,otherunitarity2,spurion,Nick,Otherhiggsless,CuringIlls,MSUdeloc,otherdelocalizedfermions}.

\subsection{Large energy behavior of scattering amplitudes}

\begin{figure}[ht]
\centerline{\includegraphics[width=0.4\hsize]{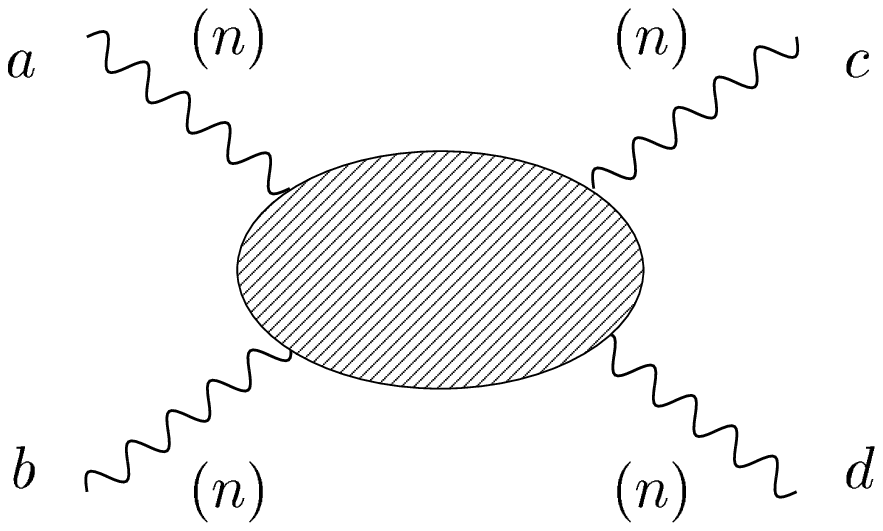}}
\caption[]{Elastic scattering of longitudinal modes of KK gauge bosons,  $n+n\to n+n$,
with the gauge index structure $a+b\to c+d$.}
\label{fig:scattering}
\end{figure}

\begin{figure}[ht]
\centerline{\includegraphics[width=0.4\hsize]{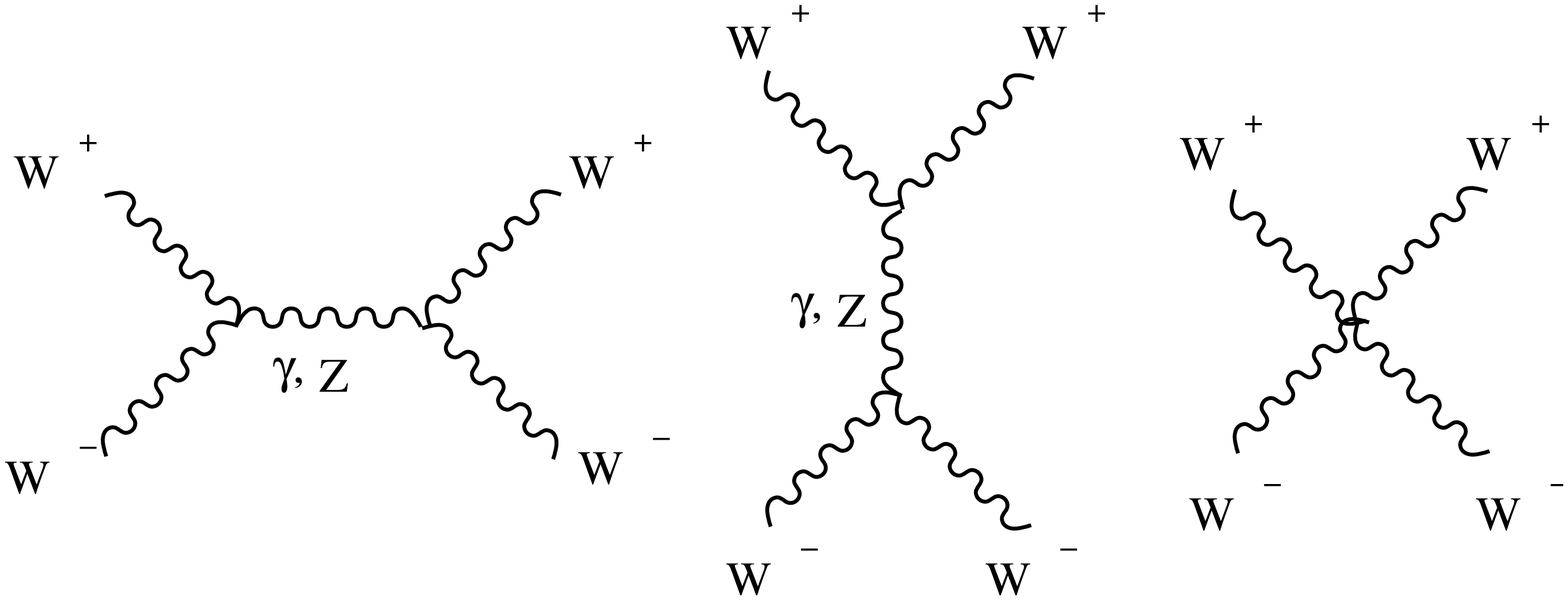}}
\caption[]{The tree-level diagrams contributing to the scattering
of massive longitudinal gauge bosons in the SM without a Higgs.}
\label{fig:SMscattering}
\end{figure}

\begin{figure}[!ht]
\centerline{\includegraphics[width=0.5\hsize]{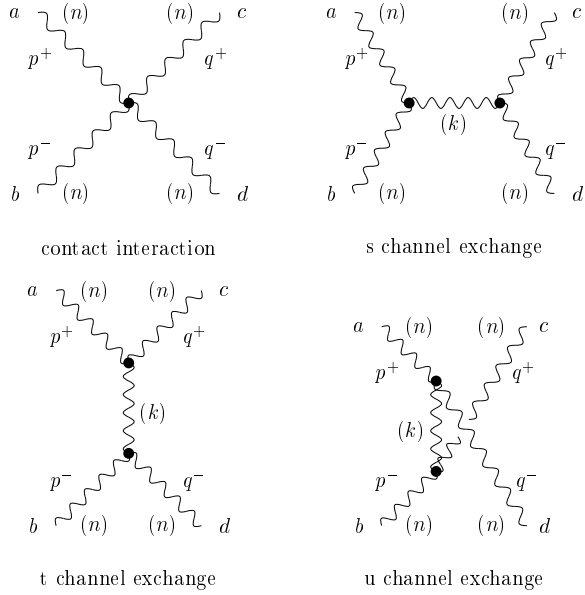}}
 \caption{The four diagrams contributing at
tree level to the elastic scattering amplitude of the nth KK
mode.}\label{fig:diagrams}
\end{figure}

Our aim is to build a higgsless model of electroweak symmetry
breaking using BC breaking in extra dimensions. However, usually
there is a problem in theories with massive gauge bosons without a
higgs scalar: the scattering amplitude of longitudinal gauge
bosons will grow with the energy and violate unitarity at a low
scale. What we would like to first understand is what happens to
this unitarity bound in a theory with extra dimensions. For
simplicity we will be focusing on the elastic scattering of the
longitudinal modes of the n$^{th}$ KK mode (see
fig.~\ref{fig:scattering}). The kinematics of this process are
determined by the longitudinal polarization vectors and the
incoming and outgoing momenta:

\begin{eqnarray}
&& \epsilon_\mu =(\frac{|\vec{p}|}{M},\frac{E}{M} \frac{\vec{p}}{|\vec{p}|})
\nonumber \\
&& p^{in}_\mu=(E,0,0,\pm \sqrt{E^2-M_n^2}) \nonumber \\
&& p_\mu^{out}=(E,\pm\sqrt{E^2-M_n^2} \sin \theta,0,
\pm\sqrt{E^2-M_n^2} \cos \theta ).
\end{eqnarray}
The diagrams that can contribute to this scattering amplitude in a
theory with massive gauge bosons (but no scalar Higgs) are given
in Fig.~\ref{fig:SMscattering} (where the $E$-dependence can be
estimated from $\epsilon \sim E, p_\mu \sim E$ and a propagator
$\sim E^{-2}$). This way we find that the amplitude could grow as
quickly as $E^4$, and then for $E\gg M_W$ can expand the amplitude
in decreasing powers of $E$ as
 \begin{equation}
\mathcal{A}= A^{(4)} \frac{E^4}{M_n^4} +
A^{(2)} \frac{E^2}{M_n^2}+
A^{(0)}+{\cal O}\left(\frac{M_n^2}{E^2}\right).
 \end{equation}
In the SM (and any theory where the gauge kinetic terms form the
gauge invariant combination $F_{\mu\nu}^2$) the $A^{(4)}$ term
automatically vanishes, while $A^{(2)}$ is only cancelled after taking the
Higgs exchange diagrams into account.

In the case of a theory with an extra dimension with BC breaking
of the gauge symmetry there are no Higgs exchange diagrams,
however one needs to sum up the exchanges of all KK modes, as in
Fig.~\ref{fig:diagrams}. As a result we will find the following
expression for the terms in the amplitudes that grow with energy:
\begin{eqnarray}
A^{(4)}= i \left(  g^2_{nnnn} - \sum_k  g_{nnk}^2  \right)
\left( f^{abe} f^{cde}  (3+6 \cos \theta - \cos^2 \theta)
 +2   (3- \cos^2 \theta)
 f^{ace} f^{bde}          \right),
\label{E4}
\end{eqnarray}
In order for the term $A^{(4)}$ to vanish it is enough to ensure
that the following sum rule among the coupling of the various KK
modes is satisfied:
\begin{equation}
g^2_{nnnn} = \sum_k  g_{nnk}^2.\label{A4sumrul}
\end{equation}
Assuming $A^{(4)}=0$ we get
\begin{eqnarray}
        \label{E2simple}
A^{(2)}=
\frac{i}{M_{n}^2}
\left(
4    g_{nnnn} M_{n}^2
-3 \sum_k  g_{nnk}^2     M_{k}^2
\right)
\left(
f^{ace} f^{bde}  - \sin^2 \sfrac{\theta}{2} \ f^{abe} f^{cde}
\right) .
\end{eqnarray}
Here $g_{nnnn}^2$ is the quartic self-coupling of the n$^{th}$ massive
gauge field, while $g_{nnk}$ is the cubic coupling between the KK modes.
In theories with extra dimensions these are of course related to the
extra dimensional wave functions $f_n(y)$ of the various modes as
\begin{eqnarray}
&& g_{mnk}=g_5 \int dy f_m (y) f_n (y) f_k (y),
\nonumber \\
\label{eq:quartic}
&& g^{2}_{mnkl}=g_5^2  \int dy f_m (y) f_n (y) f_k (y) f_l (y).
\end{eqnarray}
The most important point about the amplitudes in
(\ref{E4}-\ref{E2simple}) is that they only depend on an overall
kinematic factor multiplied by an overall expression of the
couplings. Assuming that the relation (\ref{A4sumrul}) holds we
can find a sum rule that ensures the vanishing of the $A^{(2)}$
term:
\begin{equation}
    g_{nnnn} M_{n}^2
=\frac{3}{4} \sum_k  g_{nnk}^2     M_{k}^2
\end{equation}

Amazingly, higher dimensional gauge invariance will ensure that
both of these sum rules are satisfied as long as the breaking of
the gauge symmetry is spontaneous. For example, it is easy to show
the first sum rule via the completeness of the wave functions
$f_n(y)$:
\begin{equation}
        \label{E4cond}
\int_0^{\pi R} dy\, f_n^4(y) = \sum_k \int_0^{\pi R} dy \int_0^{\pi R} dz\
f_n^2(y)f_n^2(z) f_k^{\vphantom{2}}(y) f_k^{\vphantom{2}}(z),
\end{equation}
and using the completeness relation
\begin{equation}
        \label{completeness}
\sum_k f_k(y) f_k(z) =\delta (y-z),
\end{equation}
we can see that the two sides will indeed agree. One can similarly
show that the second sum rule will also be satisfied if the
boundary conditions are natural ones (as defined in
Section~\ref{sec:lec1}) and all terms in the Lagrangian (including
boundary terms) are gauge invariant.

What we see from the above analysis is that in any gauge invariant
extra dimensional theory the terms in the amplitude that grow with
the energy will cancel. However, this will not automatically mean
that the theory itself is unitary. The reason is that there are
two additional worries: even if $A^{(4)}$ and $A^{(2)}$ vanish
$A^{(0)}$ could be too large and spoil unitarity. This is what
happens in the SM if the Higgs mass is too large. In the extra
dimensional case what this would mean is that the extra KK modes
would make the scattering amplitude flatten out to a constant
value.  However if the KK modes themselves are too heavy then this
flattening out will happen too late when the amplitude already
violates unitarity. The other issue is that in a theory with extra
dimensions there are infinitely many KK modes and thus as the
scattering energy grows one should not only worry about the
elastic channel, but the ever growing number of possible inelastic
final states. The full analysis taking into account both effects
has been performed in~\cite{Papucci}, where it was shown that
after taking into account the opening up of the inelastic channels
the scattering amplitude will grow linearly with energy, and will
always violate unitarity at some energy scale. This is a
consequence of the intrinsic non-renormalizability of the higher
dimensional gauge theory. It was found in \cite{Papucci} that the
unitarity violation scale due to the linear growth of the
scattering amplitude is equal (up to a small numerical factor of
order $2-4$) to the cutoff scale of the 5D theory obtained from
naive dimensional analysis (NDA). This cutoff scale can be
estimated in the following way. The one-loop amplitude in 5D is
proportional to the 5D loop factor
\begin{equation}
\frac{g_5^2}{24 \pi^3}.
\end{equation}
The dimensionless quantity obtained from this loop factor is
\begin{equation}
\frac{g_5^2 E}{24 \pi^3},
\end{equation}
where $E$ is the scattering energy. The cutoff scale can be
obtained by calculating the energy scale at which this loop factor
will become order one (that is the scale at which the loop and
tree-level contributions become comparable). From this we get
\begin{equation}
\Lambda_{NDA}=\frac{24 \pi^3}{g_5^2}.\label{NDA}
\end{equation}
We can express this scale using the matching of the higher dimensional
and the lower dimensional gauge couplings. In the simplest theories
this is usually given by
\begin{equation}
g_5^2= \pi R g_4^2,
\end{equation}
where $\pi R$ is the length of the interval, and $g_4$ is the effective 4D
gauge coupling.
So the final expression of the cutoff scale can be given as
\begin{equation}
\Lambda_{NDA}=\frac{24 \pi^2}{g_4^2 R}.
\end{equation}
We will see that in the Higgsless models $1/R$ will be replaced by
$M_W^2/M_{KK}$, where $M_W$ is the physical W mass, and $M_{KK}$
is the mass of the first KK mode beyond the W. Thus the cutoff
scale will indeed be lower if the mass of the KK mode used for
unitarization is higher. However, this $\Lambda_{NDA}$ could be
significantly higher than the cutoff scale in the SM without a
Higgs, which is around 1.8 TeV. We will come back to a more
detailed discussion of $\Lambda_{NDA}$ in higgsless models at the
end of this section.

\subsection{Naive Higgsless toy model}

Now that we have convinced ourselves that one can use KK gauge
bosons to delay the unitarity violation scale basically up to the
cutoff scale of the higher dimensional gauge theory, we should
start looking for a model that actually has these properties and
resembles the SM. It should have a massless photon, a massive
charged gauge boson to be identified with the $W$ and a somewhat
heavier neutral gauge boson to be identified with the $Z$. Most
importantly, we need to have the correct SM mass ratio (at
tree-level)
\begin{equation}
\frac{M_W^2}{M_Z^2} =\cos^2 \theta_W =\frac{g^2}{g^2+g'^2},
\end{equation}
where $g$ is the SU(2)$_L$ gauge coupling and $g'$ the U(1)$_Y$ gauge coupling
of the SM. We would like to use BC's to achieve this. This seems to be very
hard at first sight, since we need to somehow get a theory where the masses of
the KK modes are related to the gauge couplings. Usually the KK masses
are simply integer or half-integer multiples of $1/R$. For example, if we
look at a very naive toy model with an SU(2) gauge group in the bulk,
we could consider the following BC's for the various gauge directions:
\begin{eqnarray}
&& \partial_y A^3_\mu =0 \ {\rm at} \ y=0, \pi R, \nonumber \\
&& \partial_y A^{1,2}_\mu =0 \ {\rm at} \ y=0 \nonumber \\
&& A^{1,2}_\mu =0 \ {\rm at} \ y=\pi R.
\end{eqnarray}
With these BC's the wave functions for the various gauge fields will be
for $A^3$
\begin{equation}
f^{(n)}_3 (y)=\cos \frac{ny}{R}
\end{equation}
while for the $1,2$ directions
\begin{equation}
 f^{(m)}_{1,2} (y)=\cos \frac{(2m+1)y}{2 R}
\end{equation}
The mass spectrum then is
\begin{eqnarray}
&& A^3 \to m_n =\frac{n}{R}, \ \ 0, \frac{1}{R}, \frac{2}{R}, \ldots
\nonumber \\
&& A^{1,2} \to m_m =\frac{m+\frac{1}{2}}{R}, \ \ \frac{1}{2R}, \frac{3}{2R},
\ldots
\end{eqnarray}
This spectrum somewhat resembles that of the SM in the sense that
there is a massless gauge boson that can be identified with the
$\gamma$, a pair of charged massive gauge bosons that can be
identified with the $W^{\pm}$, and a massive neutral gauge boson
that can be identified with the $Z$. However, we can see that the
mass ratio of the W and Z is given by
\begin{equation}
\frac{M_Z}{M_W}=2,
\end{equation}
and another problem is that the first KK mode of the W,Z is given by
\begin{equation}
\frac{M_{Z'}}{M_Z}= 2, \ \ \frac{M_{W'}}{M_W}= 3.
\end{equation}
Thus, besides getting the totally wrong W/Z mass ratio there would
also be additional KK states at masses of order 250 GeV, which is
phenomenologically unacceptable. We will see that both of these
problems can be resolved by going to a warped higgsless model with
custodial SU(2).

\subsection{Custodial SU(2) and flat space higgsless model}

We have seen above that a major question in building a realistic
higgsless model is how to ensure that the W/Z mass ratio agrees
with the tree-level result. Let us first understand why the tree
result in the SM is given by
\begin{equation}
\rho \equiv \frac{M_W^2}{\cos^2 \theta_W M_Z^2}=1.
\end{equation}
The electroweak symmetry in the SM is broken by the Higgs scalar
$H$, which transforms as a ${\bf 2}_{\frac{1}{2}}$ under
SU(2)$_L\times$ U(1)$_Y$. The Higgs potential is given by
\begin{equation}
V(H)=-\mu^2 H^\dagger H+\lambda (H^\dagger H)^2.
\end{equation}
This potential is only a function of $H^\dagger H$, which can also be written
as
\begin{equation}
H^\dagger H= h_1^2+h_2^2+h_3^2+h_4^2,
\label{custodial}
\end{equation}
where the Higgs doublet has been written in terms of its real and imaginary
components as
\begin{equation}
H=\left( \begin{array}{c} h_1+i h_2 \\ h_3 +i h_4
\end{array} \right).
\end{equation}
We can see from (\ref{custodial}) that the Higgs potential
actually has a bigger global symmetry than SU(2)$_L\times$
U(1)$_Y$: it is invariant under the full SO(4) rotation of the
four independent real fields in the Higgs doublet. The SO(4) group
is actually not a simple group, but rather equivalent to
SU(2)$_L\times$SU(2)$_R$. The origin of the SU(2)$_R$ symmetry can
also be understood as follows. A doublet of SU(2) is a pseudo-real
representation, which means that the complex conjugate of the
doublet is equivalent to the doublet itself. The way this
manifests itself is if we consider the doublet to be a field with
a lower SU(2) index $H_i$, then the complex conjugate would
automatically have an upper index. However, using the SU(2)
epsilon $\epsilon_{ij}$ this can be lowered again, and so $H_i$
and $i \epsilon_{ij} (H^*)^j$ transform in the same way. This
means that in addition to an SU(2)$_L$ acting on the usual SU(2)
index, there is another SU(2) symmetry that mixes $H$ with
$\epsilon H^*$. To make this more intuitive, we could write a 2 by
2 matrix as
\begin{equation}
\left( \begin{array}{cc} H & \epsilon H^* \end{array}\right),
\label{leftright}
\end{equation}
and then the ordinary SU(2)$_L$ would act from the left, and the additional
SU(2)$_R$ would be a global symmetry acting on this matrix from the
right.

Once the Higgs scalar gets a VEV, it will break the SO(4) global
symmetry to an SO(3) subgroup. In the SU(2) language this means
that SU(2)$_L\times$SU(2)$_R$ is broken to the diagonal subgroup
$SU(2)_D$. This is most easily seen from the representation in
(\ref{leftright}) since then we have a matrix whose VEV is given
by ${\rm diag}(v,v)$, and obviously leaves the diagonal subgroup
unbroken.

The claim is that once such an SU(2)$_D$ subgroup (which is
usually referred to as the custodial SU(2) symmetry of the Higgs
potential) is left unbroken during electroweak symmetry breaking,
it is guaranteed that the $\rho$-parameter will come out to be one
at tree level. Let us quickly prove that this is indeed the case.
The generic description of the global symmetry breaking pattern
SU(2)$_L\times$ SU(2)$_R \to$ SU(2)$_D$ can be achieved via the
non-linear $\sigma$-model which will describe the physics of the 3
Goldstone-modes appearing in this symmetry breaking. In this
description the possible massive Higgs modes are integrated out.
This model will give all the consequences of the global
symmetries. In this model the Goldstone fields are represented by
a 2 by 2 unitary matrix $\Sigma$, which is given in terms of the
Goldstone modes $\pi^a$ as
\begin{equation}
\Sigma =e^{i \frac{\pi^a \tau^a}{f}},
\end{equation}
and transforms under SU(2)$_L\times$SU(2)$_R$ as $\Sigma \to U_L
\Sigma U_R^\dagger$. One can think of the SU(2)$_L\times$ U(1)$_Y$
electroweak symmetries as a subgroup of SU(2)$_L\times$ SU(2)$_R$,
with U(1)$_Y\subset$ SU(2)$_R$. This gauging of a subgroup of the
global symmetries will explicitly break some of the global
symmetries, but this is easily incorporated into the non-linear
$\sigma$-model description. (Note, that in the presence of
fermions the global symmetry needs to be enlarged to
SU(2)$_L\times$ SU(2)$_R\times$ U(1)$_{B-L}$, and then
U(1)$_Y\subset$ SU(2)$_R\times $U(1)$_{B-L}$, where U(1)$_{B-L}$
is the baryon number minus lepton number symmetry.) The covariant
derivative is then given by
\begin{equation}
D_\mu=\partial_\mu -i g \frac{\tau^a}{2} A_\mu^a -i\frac{g'}{2} B_\mu.
\end{equation}
The leading kinetic term for the pions is given by
\begin{equation}
f^2 {\rm Tr} (D_\mu \Sigma^\dagger)(D^\mu \Sigma ).
\end{equation}
Expanding this expression in the pion fluctuations will result in
mass terms for the gauge fields
\begin{equation}
M_W^2=\frac{g^2 f^2}{4}, \ \ M_Z^2=\frac{(g^2+g'^2) f^2}{4},
\end{equation}
and thus
\begin{equation}
\frac{M_W^2}{M_Z^2}=\frac{g^2}{g^2+g'^2}.
\end{equation}
Note, that in the above derivation the only information that has been used
was the global symmetry breaking pattern SU(2)$_L\times$SU(2)$_R\to
$ SU(2)$_D$, with the appropriate subgroup being gauged. Once this
symmetry breaking pattern is established, it is guaranteed that the tree-level
prediction for the $\rho$-parameter will be 1.

\begin{figure}[ht]
\centerline{\includegraphics[width=0.6\hsize]{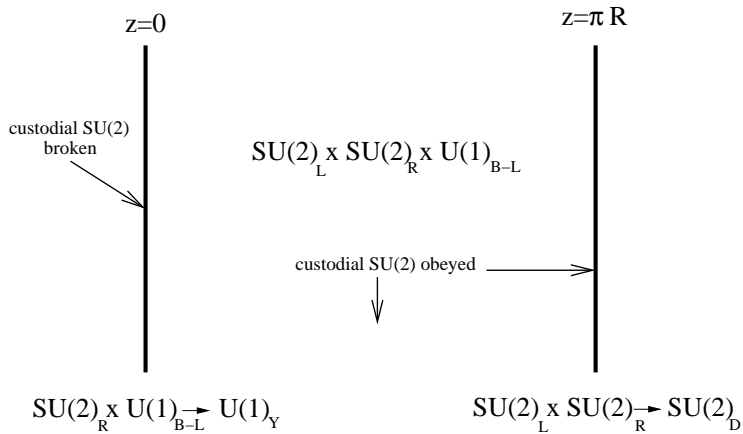}}
\caption[]{The symmetry breaking structure of the flat space
higgsless toy model.} \label{flathiggsless}
\end{figure}

From the above discussion it is clear that in order to find a
higgsless model with the correct W/Z mass ratio one needs to find
an extra dimensional model that has the custodial SU(2) symmetry
incorporated~\cite{ADMS}. Once such a construction is found, the
gauge boson mass ratio will automatically be the right one.
Therefore we need to somehow involve SU(2)$_R$ in the
construction. The simplest possibility is to put an entire
SU(2)$_L\times$SU(2)$_R\times$U(1)$_{B-L}$ gauge group in the bulk
of an extra dimension~\cite{CGMPT}. In order to mimic the symmetry
breaking pattern in the SM most closely, we assume that on one of
the branes the symmetry breaking is SU(2)$_L\times$SU(2)$_R\to$
SU(2)$_D$, with $U(1)_{B-L}$ unbroken. On the other boundary one
needs to reduce the bulk gauge symmetry to that of the SM, and
thus have a symmetry breaking pattern
SU(2)$_R\times$U(1)$_{B-L}\to $ U(1)$_Y$, which is illustrated in
Fig.~\ref{flathiggsless}.

We denote by $A^{R\,a}_{M}$, $A^{L\,a}_{M}$ and $B_M$ the gauge
bosons of $SU(2)_R$, $SU(2)_L$ and $U(1)_{B-L}$ respectively;
$g_{5L}$ and $g_{5R}$ are the the gauge couplings of the two
$SU(2)$'s and $\tilde{g}_5$, the gauge coupling of the
$U(1)_{B-L}$.  In order to obtain the desired BC's as discussed
above we need to follow the procedure laid out in the first
lecture. We assume that there is a boundary Higgs on the left
brane in the representation $(1,2)_{\frac{1}{2}}$ under
SU(2)$_L\times$SU(2)$_R\times$U(1)$_{B-L}$, which will break
SU(2)$_R\times$U(1)$_{B-L}$ to U(1)$_Y$. We could also use the
more conventional triplet representation under SU(2)$_R$ which
will allow us to get neutrino masses later on. On the right brane
we assume that there is a bi-doublet higgs in the representation
$(2,2)_0$ which breaks the electroweak symmetry as in the SM:
SU(2)$_L\times$SU(2)$_R \to$SU(2)$_D$. We will then take all the
Higgs VEV's to infinity in order to decouple the boundary scalars
from the theory, and impose the natural boundary conditions as
described in the first lecture. The BC's we will arrive at then
are:
\begin{eqnarray}
&
{\rm at }\  y=0:
&
\left\{
\begin{array}{l}
\partial_z (g_{5R}B_\mu+\tilde{g}_5 A_\mu^{R3})=0 \
,\partial_z A^{La}_\mu  = 0, \ A_\mu^{R1,2} = 0,
\\
\tv{15}
\tilde{g}_5 B_\mu -g_{5R} A_\mu^{R3} = 0,
\end{array}
\right.
\label{bc1}\\
&
{\rm at }\ y=\pi R:
&
\left\{
\begin{array}{l}
\partial_z (g_{5R} A^{L\,a}_\mu+g_{5L} A^{R\,a}_\mu)=0, \
\partial_z B_\mu =0, \ g_{5L}A^{L\,a}_\mu-g_{5R}A^{R\,a}_\mu=0.
\end{array}\right.
\label{bc2}
\end{eqnarray}
The BC's for the $A_5$ and $B_5$ components will be the opposite
of the 4D gauge fields as usual, i.e. all Dirichlet conditions
should be replaced by Neumann and vice versa. The next step to
determine the mass spectrum is to find the right KK decomposition
of this model. First of all, none of the $A_5$ and $B_5$
components have a flat BC on both ends. This means that there will
be no zero mode in these fields, and as we have seen all the
massive scalars are unphysical, since they are just gauge
artifacts (supplying the longitudinal components of the massive KK
towers). So we will not need to discuss the modes in these fields.
The main point to observe about the KK decomposition of the gauge
fields is that the BC's will mix up the states in the various
components. This will imply that a single 4D mode will live in
several different 5D fields. Since in the bulk there is no mixing,
and we are discussing at the moment a flat 5D background, the wave
functions will be of the form $f_k(z)\propto a \cos M_k z+b \sin
M_k z$. If we make the simplifying assumption that
$g_{5L}=g_{5R}$, then the KK decomposition will be somewhat
simpler than the most generic one, and given by (we denote by
$A^{L,R\, \pm}_\mu$ the linear combinations $\sfrac{1}{\sqrt{2}}
(A^{L,R\, 1}\mp i A^{L,R\, 2})$):
\begin{eqnarray}
        \label{eq:KKB1}
B_\mu (x,y) & = & g\,  a_0 \gamma_\mu (x)
+  {g'} \sum_{k=1}^{\infty} b_k \cos (M^{Z}_k y) \, Z^{(k)}_\mu (x)\, ,
\\
        \label{eq:KKAL3}
A^{L\, 3}_\mu (x,y) & = &
{g'} \, a_0 \gamma_\mu (x)
-  g \sum_{k=1}^{\infty}   b_k  \frac{\cos (M^{Z}_k (y-\pi R))}{2 \cos (M^{Z}_k \pi R)} \, Z^{(k)}_\mu (x) \, ,
\\
        \label{eq:KKAR3}
A^{R\, 3}_\mu (x,y) & = &
{g'}  a_0 \gamma_\mu (x)
- g  \sum_{k=1}^{\infty}  b_k \frac{\cos (M^{Z}_k (y+\pi R))}{2 \cos (M^{Z}_k \pi R)} \, Z^{(k)}_\mu (x) \, ,
\\
        \label{eq:KKALpma}
A^{L\, \pm}_\mu (x,y) & = &
\sum_{k=1}^{\infty}   c_k \cos (M^{W}_k (y-\pi R))\, W^{(k)\, \pm}_\mu (x) \, ,
\\
        \label{eq:KKARpma}
A^{R\, \pm}_\mu (x,y) & = &
\sum_{k=1}^{\infty} c_k \cos (M^{W}_k (y+\pi R))\, W^{(k)\, \pm}_\mu (x)  \, .
\end{eqnarray}

The coefficients and the masses are then determined by imposing the BC's
on this KK decomposition. The resulting mass spectrum that we find is the
following.
The spectrum is made up of a massless photon, the gauge boson associated with
the unbroken
$U(1)_Q$  symmetry, and some towers of massive charged and neutral gauge
bosons,
$W^{(k)}$ and $Z^{(k)}$ respectively.
The masses of the $W^\pm$'s are given by
\begin{equation}
M^W_k = \frac{2k-1}{4R}, \ \ k=1,2\ldots
\end{equation}
while for $Z$'s there are two towers of neutral
gauge bosons with masses
\begin{eqnarray}
M^Z_k = \left(M_0 + \frac{k-1}{R}\right), \ \   k=1,2\ldots
\\
M^{Z'}_k = \left(-M_0 + \frac{k}{R}\right), \ \   k=1,2\ldots
\end{eqnarray}
where $M_0=\sfrac{1}{\pi R} \arctan \sqrt{1+2 {{g'}}^2/g^2}$. Note that
$1/(4R)<M_0<1/(2R)$ and thus the $Z'$'s are heavier than the $Z$'s
($M^{Z'}_k>M^Z_k$). We also get that the lightest $Z$ is heavier than
the lightest $W$ ($M^Z_1>M^W_1$), in agreement with the SM spectrum. However,
the mass ratio of W/Z is given by
\begin{equation}
\frac{M_W^2}{M_Z^2}=\frac{\pi^2}{16}\arctan^{-2}
\sqrt{1+\frac{g_{4D}^{' 2}}{g_{4D}^2}}
\sim 0.85,
\end{equation}
and hence the $\rho$ parameter is
\begin{equation}
\rho=\frac{M_W^2}{M_Z^2 \cos^2 \theta_W}\sim 1.10 \ .
\end{equation}
Thus the mass ratio is close to the SM value, however the ten
percent deviation is still huge compared to the experimental
precision. The reason for this deviation is that while the bulk
and the right brane are symmetric under custodial SU(2), the left
brane is not, and the KK wave functions do have a significant
component around the left brane, which will give rise to the large
deviation from $\rho =1$. Thus one needs to find a way of making
sure that the KK modes of the gauge fields do not very much feel
that left brane, but are repelled from there, and only the
lightest (almost zero modes) $\gamma ,Z,W^{\pm}$ will have a large
overlap with the left brane.

\subsection{The AdS/CFT correspondence}

We have seen above that one would need to modify the flat space
setup such that the KK modes get pushed away from the left brane
without breaking any new symmetries. There are two possible ways
that this can be done (and in fact we will see that these two are
basically equivalent to each other). One possibility is to simply
add a large brane kinetic term for the gauge fields on the left
brane where custodial SU(2) is violated~\cite{BPR}. The effect of this will
be exactly to push away the heavy KK modes (since as we have seen
in the first lecture the BC does depend on the eigenvalue). This
way the custodial SU(2) can be approximately restored in the KK
sector of the theory. The second possibility which we will be
pursuing here is to use the AdS/CFT correspondence. This has been first pointed out
in~\cite{ADMS}.

It has been realized by Maldacena~\cite{Maldacena} in 1997 that certain string
theories on an anti-de Sitter (AdS) background are actually
equivalent to some 4D conformal field theories. The crucial
ingredient was that the conformal group SO(2,4) is equivalent to
the isometries of 5D AdS space, whose metric is given by
\begin{equation}
ds^2=  \left( \frac {R}{z} \right)^2   \Big( \eta_{\mu \nu} dx^\mu dx^\nu - dz^2 \Big)
\end{equation}
where $0<z<\infty$ is the radial AdS coordinate. Basically the
point is that besides the usual 5D Poincare transformations this
metric has an additional rescaling invariance $z\to \alpha z,
x_\mu \to \alpha x_\mu$. Using several checks Maldacena was able
to gather convincing evidence that ${\cal N}=4$ supersymmetric
Yang Mills theory in a certain limit is equivalent to type IIB
string theory on AdS$_5\times S^5$. This field theory is
automatically conformal due to the number of supercharges. One
crucial observation is that the coordinate along the AdS direction
actually corresponds to an energy scale in the CFT. This is quite
clear from the above mentioned rescaling invariance. Rescaling $z$
implies rescaling $x$. But rescaling $x$ means changing the energy
scale in the CFT. So this will imply that the $z\to 0$ region
corresponds to the most energetic sector of the CFT $E\to \infty$,
while the $z\to \infty$ corresponds to low energies $E\to 0$. The
other important observation of this equivalence is that the field
theory has a large global symmetry SU(4)$_R$. The way this is
realized in the string theory side is that SO(6)$=$SU(4) is the
isometry of the 5D sphere $S^5$.  Thus there will be massless
gauge boson corresponding to this SU(4) global symmetry in the
AdS$_5$ theory. What this seems to suggest is that the AdS$_5$
bulk is really supplying the modes of the CFT itself, while the
global symmetries of the CFT will manifest themselves in gauge
fields appearing in the 5D theory. One final crucial ingredient
needed for us is how this correspondence will be modified if the
AdS space is not infinite, but we are considering only a finite
interval (slice) of 5D AdS space. In this case we clearly do not
have the full conformal invariance, since the appearance of the
boundaries of the slice of AdS will explicitly break it. One way
of interpreting the appearance of a boundary close to $z=0$
(usually called the UV brane or Planck brane since it corresponds
to high energies) is that the field theory has an explicit cutoff
corresponding to this energy scale. If the cutoff is at $z=R$, the
field theory will have a UV cutoff $\Lambda =\frac{1}{R}$. The
interpretation of the other boundary (usually referred to as IR
brane or TeV brane) is trickier. It has been argued
in~\cite{APR,RZ} that the proper interpretation of such an IR
brane is that the CFT {\it spontaneously} breaks the conformal
invariance at low energies. The location of the IR brane will
supply an IR cutoff. For those familiar with the phenomenology of
the Randall-Sundrum models, this can be explained by realizing
that the KK spectrum of the fields will be localized on this IR
brane. In the presence of the IR brane there will also be a
discrete spectrum for these KK modes. These discrete KK modes can
be thought of as the composites (bound states) formed by the CFT
after it breaks conformality and becomes strongly interacting
(confining). The analogy could be a theory that is very slowly
running (its $\beta$-function is very close to zero), then after a
long period of slow running (``walking'') the theory will become
strongly interacting, and the theory will confine (if it is
QCD-like) and form bound states. What happens then to the gauge
fields if they are in the bulk of a finite slice of the AdS space?
It depends on their BC's. The main difference between the full AdS
and the case of a slice is that while a gauge field zero mode is
not normalizable in an infinite AdS space, it will become
normalizable in the case of a slice. So this means that if there
is a zero mode present, then one would need to identify this as a
weakly gauged global symmetry of the CFT. Whether the gauge field
in the finite slice actually has a zero mode or not, depends on
its BC's. If it has Dirichlet BC on the UV brane, then the zero
mode will pick up a mass of the order of the scale at the UV brane
(that is proportional to the UV cutoff), so it will be totally
eliminated from the theory. Thus in this case even in a finite
slice the symmetry should be thought of as a global symmetry only.
However, if the BC on the UV brane is Neumann, while on the IR
brane is Dirichlet, then the zero mode picks up a mass of the
order of the IR cutoff (the confinement scale, the KK scale of the
other resonances), so the way this should be interpreted is that
the CFT became strongly interacting, and that breaking of
conformality also resulted in breaking the weakly gauged global
symmetry spontaneously. Later on it was also realized that perhaps
supersymmetry may not be necessary for such a correspondence to
exist. So let us summarize the rules laid out above for the
AdS/CFT correspondence (at least the ones relevant for model
building) in Table~\ref{adscfttab}.
\begin{table}[h]
\begin{center}
\begin{tabular}{lll}
 Bulk of AdS \hspace*{4cm}& $\leftrightarrow$ & \hspace*{1cm}CFT \\[.25cm]
 Coordinate ($z$) along AdS & $\leftrightarrow$  & \hspace*{1cm}Energy scale in CFT \\[.25cm]
 Appearance of UV brane & $\leftrightarrow$ & \hspace*{1cm}CFT has a cutoff \\[.25cm]
 Appearance of IR brane & $\leftrightarrow$ &
\hspace*{1cm}\begin{minipage}{6cm} conformal symmetry broken
spontaneously by CFT \end{minipage}\\[.35cm]
 KK modes localized on IR brane & $\leftrightarrow$ & \hspace*{1cm}composites of CFT \\[.25cm]
 Modes on the UV brane & $\leftrightarrow$ & \hspace*{1cm}Elementary fields
coupled
to CFT \\[.25cm]
 Gauge fields in bulk & $\leftrightarrow$ & \hspace*{1cm}CFT has a global symmetry  \\[.25cm]
\begin{minipage}{6.5cm} Bulk gauge symmetry broken\\ on UV brane\end{minipage} & $\leftrightarrow$ & \hspace*{1cm}Global symmetry not gauged \\[.4cm]
\begin{minipage}{6.5cm} Bulk gauge symmetry unbroken\\ on UV brane\end{minipage} & $\leftrightarrow$ &
\hspace*{1cm}Global
symmetry weakly gauged \\[.35cm]
 Higgs on IR brane & $\leftrightarrow$ &
\hspace*{1cm}\begin{minipage}{6cm}CFT becoming strong produces
composite Higgs \end{minipage}\\[.35cm]
 \begin{minipage}{6cm}Bulk gauge symmetry broken\\ on IR brane by
BC's \end{minipage}& $\leftrightarrow$ &
\hspace*{1cm}\begin{minipage}{6cm}Strong dynamics that breaks CFT also breaks gauge symmetry \end{minipage}\\
\end{tabular}
\end{center}
\caption{Relevant rules for model building using the AdS/CFT
correspondence}\label{adscfttab}
\end{table}

\begin{figure}[h]
\centerline{\includegraphics[width=0.6\hsize]{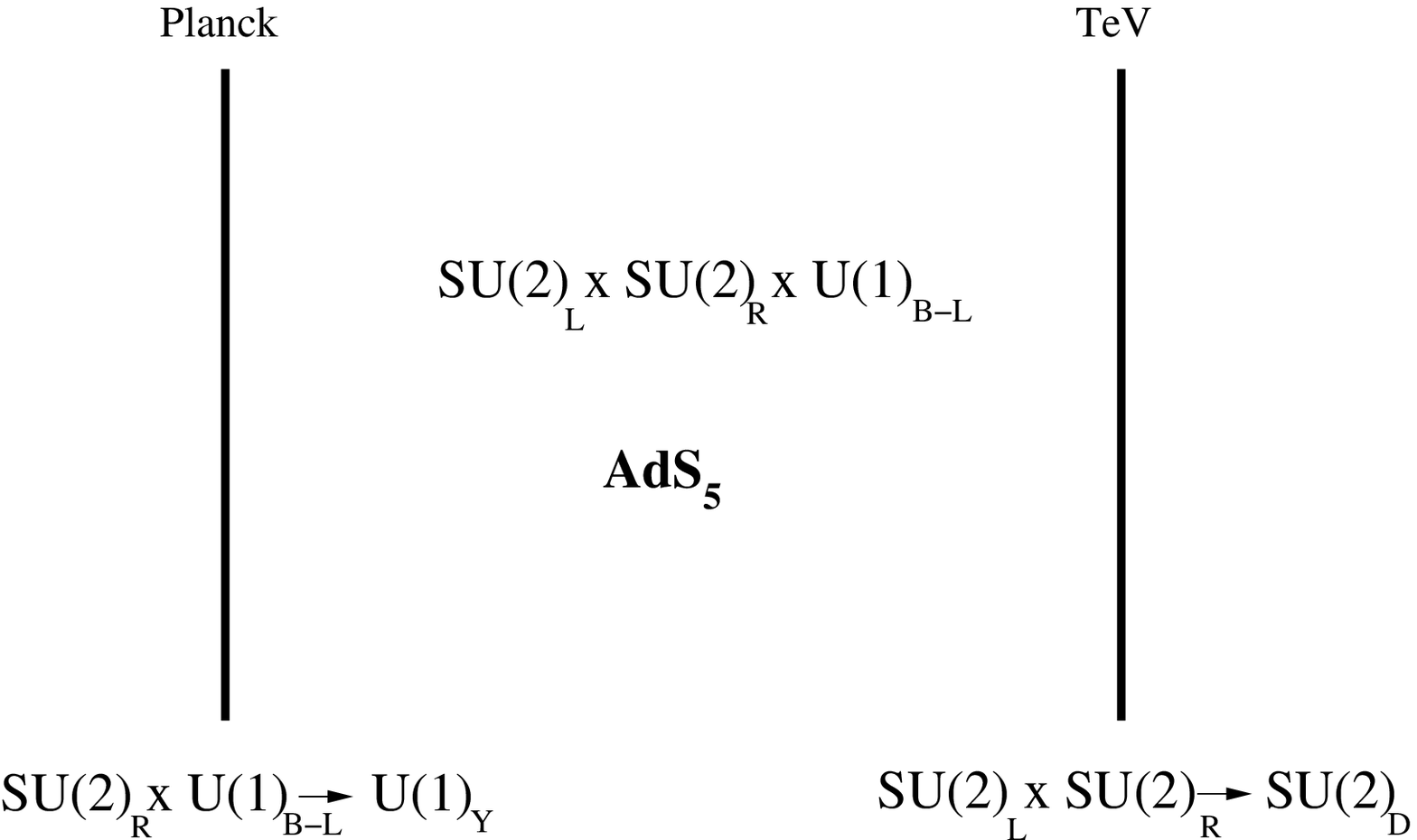}}
\caption[]{The symmetry breaking structure of the warped higgsless
model.} \label{fig:higgsless}
\end{figure}

Using the rules of the correspondence found in
Table~\ref{adscfttab} we can now relatively easily find the theory
that we are after. We want a theory that has an $SU(2)_L\times
SU(2)_R\times U(1)_{B-L}$ global symmetry, with the $SU(2)_L\times
U(1)_Y$ subgroup weakly gauged, and broken by BC's on the IR
brane. To have the full global symmetry, we need to take
$SU(2)_L\times SU(2)_R\times U(1)_{B-L}$ in the bulk of AdS$_5$.
To make sure that we do not get unwanted gauge fields at low
energies, we need to break $SU(2)_R\times U(1)_{B-L}$ to $U(1)_Y$
on the UV brane, which we will do by BC's as in the flat case.
Finally, the boundary conditions on the TeV brane break
$SU(2)_L\times SU(2)_R$ to $SU(2)_D$, thus providing for
electroweak symmetry breaking. This setup is illustrated in
Fig.~\ref{fig:higgsless}. Note, that it is practically identical
to the flat space toy model considered before, except that the
theory is in AdS space.

\subsection{The warped space higgsless model}

Before coming to the detailed prediction of the mass spectrum of
the warped space higgsless model outlined above, let us first
briefly discuss how to deal with a gauge theory in an AdS
background~\cite{RSbulk,RandallSchwartz}. We will be considering a
5D gauge theory in the fixed gravitational background
\begin{equation}
ds^2=  \left( \frac {R}{z} \right)^2   \Big( \eta_{\mu \nu} dx^\mu dx^\nu - dz^2 \Big)
\label{adsmetric}
\end{equation}
where $z$ is on the
    interval $[R,R^\prime]$.  We will not be considering gravitational
fluctuations, that is we
are assuming that the Planck scale is sent to infinity,
while the background is frozen to be the one given above.
In RS-type models $R$ is typically
$\sim 1/M_{Pl}$ and
    $R^\prime \sim  {\rm TeV}^{-1}$. For higgsless models we will see later
on what the optimal choice for these scales are. The action for a
gauge theory on a fixed background will be given by
\begin{equation}
S=\int d^5x \sqrt{g}\left[-\frac{1}{4}
F_{MN}^aF_{PQ}^{a}g^{MP}g^{NQ}\right]
\end{equation}
Putting in the metric (\ref{adsmetric}) we find
\begin{equation}
S=\int d^4x \int_R^{R'} \frac{R}{z} \left[ -\frac{1}{4} {F_{\mu\nu}^a}^2
-\frac{1}{2} {F_{\mu 5}^a}^2 \right].
\end{equation}
To get the right gauge fixing term, we have to repeat the procedure from
the first section. The mixing term between $A_\mu$ and $A_5$ is given by
\begin{equation}
\int d^4x \int_R^{R'} dz\frac{R}{z} \partial_\mu \partial^5 A^\mu =
-\int d^4x\int_R^{R'}  dz\partial_\mu A^\mu \partial_5 \left( \frac{R}{z} A_5
\right).
\end{equation}
In the second equality we have integrated by parts and neglected a
boundary term (which is necessary for determining the BC's,
however these will not change due to the presence of the warping
so the BC's derived for the flat space model will be applicable
here as well). This implies that the gauge fixing term necessary
in the warped case is given by
\begin{equation}
S_{gf}=-\int d^4x \int_R^{R'}dz \frac{1}{2\xi} \frac{R}{z} \left[
\partial_\mu A^\mu -\xi \frac{z}{R} \partial_5 \left(\frac{R}{z} A_5
\right)\right]^2.
\end{equation}
Due to the chosen BC's the $A_5$ fields will have no zero modes
they will all again become massive gauge artifacts and can be
eliminated in the unitary gauge. The quadratic piece of the action
for the gauge fields will be then given by
\begin{equation}
\int d^4 x\int_R^{R'} dz \frac{R}{z}\frac{1}{2} A_\mu
\left[ \left(\partial^2-\frac{z}{R} \partial_z \left(\frac{R}{z} \partial_z
\right)\right)\eta^{\mu\nu} -\left(1-\frac{1}{\xi}\right)
\partial^\mu \partial^\nu \right] A_\nu.
\end{equation}
As before, we go to $4D$ momentum space by writing $A_\mu
(x,z)=\epsilon_\mu(p) f(z) e^{ip\cdot x}$. The equation of motion
for the wave function $f(z)$ will then become ($p^2=M^2$):
\begin{equation}
\left[ -M^2 -z \partial_5 \left( \frac{1}{z} \partial_5\right)\right] f(z)=0.
\end{equation}
Equivalently it can be written as
\begin{equation}
f''-\frac{1}{z} f'+M^2 f=0.
\end{equation}
This will lead to a Bessel equation for $g(z)$ after the substitution
$f(z)=z g(z)$:
\begin{equation}
g''+\frac{1}{z}g'+(M^2-\frac{1}{z^2}) g=0,
\end{equation}
which is a Bessel equation of order 1.  The solution is of the
form
\begin{equation}
    \label{eq:Bwv2}
f(z)=z\left(A J_1(q_k z)+BY_1(q_k z)\right).
\end{equation}

The BC's corresponding to the symmetry breaking pattern discussed above
for the warped higgsless model are identical to the ones for the
flat space case~\cite{CGPT}:
\begin{eqnarray}
&
{\rm at }\  y=0:
&
\left\{
\begin{array}{l}
\partial_z (g_{5R}B_\mu+\tilde{g}_5 A_\mu^{R3})=0 \
\partial_z A^{La}_\mu  = 0, \
A_\mu^{R1,2} = 0,
\\
\tv{15}
\tilde{g}_5 B_\mu -g_{5R} A_\mu^{R3} = 0,
\end{array}
\right.
\label{bc1w}\\
&
{\rm at }\ y=\pi R:
&
\left\{
\begin{array}{l}
\partial_z (g_{5R} A^{L\,a}_\mu+g_{5L} A^{R\,a}_\mu)=0, \
\partial_z B_\mu =0, \ g_{5L}A^{L\,a}_\mu-g_{5R}A^{R\,a}_\mu=0.
\end{array}\right.
\label{bc2w}
\end{eqnarray}
Again the BC's for the $A_5$'s are the opposite to that of the
corresponding combination of the $4D$ gauge fields, and these BC's
can be thought of as arising from Higgses on each brane in the
large VEV limit. Using the above Bessel functions the KK mode
expansion is given by the solutions to this equation which are of
the form
\begin{equation}
    \label{eq:Bwv}
\psi^{(A)}_k(z)=z\left(a^{(A)}_k J_1(q_k z)+b^{(A)}_k Y_1(q_k z)\right)~,
\end{equation}
where $A$ labels the corresponding gauge boson.
Due to the mixing of the various gauge groups, the KK decomposition is
slightly complicated but it is obtained
by simply enforcing the BC's:
\begin{eqnarray}
           \label{eq:KKB}
B_\mu (x,z) & = & g_5\,  a_0 \gamma_\mu (x)
+   \sum_{k=1}^{\infty} \psi^{(B)}_k (z)  \, Z^{(k)}_\mu (x)\, ,
\\
           \label{eq:KKAL3a}
A^{L\, 3}_\mu (x,z) & = &
{\tilde g_5} \, a_0 \gamma_\mu (x)
+   \sum_{k=1}^{\infty} \psi^{(L3)}_k (z) \, Z^{(k)}_\mu (x) \, ,
\\
           \label{eq:KKAR3a}
A^{R\, 3}_\mu (x,z) & = &
{\tilde g_5}  \, a_0 \gamma_\mu (x)
+    \sum_{k=1}^{\infty}  \psi^{(R3)}_k (z) \, Z^{(k)}_\mu (x) \, ,
\\
           \label{eq:KKALpm}
A^{L\, \pm}_\mu (x,z) & = &
  \sum_{k=1}^{\infty}  \psi^{(L\pm)}_k (z) \, W^{(k)\, \pm}_\mu (x) \, ,
\\
           \label{eq:KKARpm}
A^{R\, \pm}_\mu (x,z) & = &
  \sum_{k=1}^{\infty} \psi^{(R\pm)}_k  (z) \, W^{(k)\, \pm}_\mu (x)  \, .
\end{eqnarray}
Here $\gamma(x)$ is the 4D photon, which has a flat wavefunction due to
the
unbroken $U(1)_Q$  symmetry, and $W^{(k)\, \pm}_\mu (x)$ and
$Z^{(k)}_\mu (x)$ are the KK towers of the massive $W$ and $Z$
gauge bosons, the lowest of which are supposed to correspond to the
observed $W$ and $Z$.

To leading order in $1/R$ and for
$\log \left(R^\prime/R \right) \gg1$, the lightest solution for this
equation for the mass of the $W^\pm$'s is
\begin{equation}
M_W^2 \approx \frac{1}{R^{\prime 2} \log
\left(\frac{R^\prime}{R}\right)} \, .
\end{equation}
Note, that this result  does not depend
on the 5D gauge coupling, but only on the scales $R,R'$.
Taking $R= 10^{-19}$ GeV$^{-1}$ will fix $R^\prime= 2 \cdot 10^{-3}$
GeV$^{-1}$. The lowest mass of the $Z$ tower is approximately given by
\begin{equation}
M_Z^2  = \frac{g_5^2+2 \tilde g_5^{2}}{g_5^2+ \tilde g_5^{2}}
\frac{1}{R^{\prime 2} \log \left(\frac{R^\prime}{R}\right)}  \, .
\end{equation}
If the SM fermions are localized on the Planck brane then the
leading order expression for the effective 4D couplings will be
given by (see Section~\ref{sec:EW} for more details)
\begin{eqnarray}
&& \frac{1}{g^2}=\frac{R\log \left(\frac{R^\prime}{R}\right)}{g_{5}^2}
\nonumber \\
&& \frac{1}{g'^2}=R\log \left(\frac{R^\prime}{R}\right)\left(\frac{1}{g_{5}^2}
+\frac{1}{\tilde{g}_{5}^2}\right),
\end{eqnarray}
thus the 4D Weinberg angle will be given by
\begin{equation}
\sin \theta_W = \frac{ \tilde g_5}{\sqrt{ g_5^2+2 \tilde g_5^2} }=
\frac{ g^\prime}{\sqrt{ g^2+  g^{\prime 2}} }.
\label{SM3}
\end{equation}
We can see that to leading order the SM expression for the W/Z
mass ratio is reproduced in this theory as expected. In fact the
full structure of the SM coupling is reproduced at leading order
in $1/\log (R'/R)$, which implies that at the leading log level
there is no $S$-parameter either. An $S$-parameter in this
language would have manifested itself in an overall shift of the
coupling of the Z compared to its SM value evaluated from the $W$
and $\gamma$ couplings, which are absent at this order of
approximation. The corrections to the SM relations will appear in
the next order of the log expansion. Since $\log
\left(\frac{R^\prime}{R}\right) \sim {\cal O} (10)$, this
correction could still be too large to match the precision
electroweak data. We will be discussing the issue of electroweak
precision observables in the last lecture.

The KK masses of the W (and the Z bosons as well due to custodial
SU(2) symmetry) will be given approximately by
\begin{equation}
m_{W_n}=\frac{\pi}{2} (n+\frac{1}{2}) \frac{1}{R'} , \ \ n=1,2,\ldots .
\end{equation}
We can see that the ratio between the physical W mass and the
first KK mode is given by
\begin{equation}
\frac{m_W}{m_W'}=\frac{4}{3 \pi} \frac{1}{\sqrt{\log
\left(\frac{R^\prime}{R}\right)}}.
\end{equation}
We can see that warping will achieve two desirable properties: it
will enforce custodial SU(2) and thus automatically generate the
correct W/Z mass ratio, but it will also push up the masses of the
KK resonances of the W and Z. This will imply that we can get a
theory where the W', Z' bosons are not so light that they would
already be excluded by the LEP or the Tevatron experiments.
Finally, we can return to the issue of perturbative unitarity in
these models. In the flat space case we have seen that the scale
of unitarity violation is basically given by the NDA cutoff scale
(\ref{NDA}). However, in a warped extra dimension all scales will
be dependent on the location along the extra dimension, so the
lowest cutoff scale that one has is at the IR brane given by
\begin{equation}
\Lambda_{NDA}\sim\frac{24 \pi^3}{g_5^2} \frac{R}{R'}.
\end{equation}
Using our expressions for the 4D couplings and the W and W' masses
above we can see that~\cite{Papucci,CuringIlls}
\begin{equation}
\Lambda_{NDA}\sim \frac{12 \pi^4 M_W^2}{g^2 M_{W^{(1)}}}\,.
\end{equation}

From the formula above, it is clear that the heavier the
resonance, the lower the scale where perturbative unitarity is
violated. This also gives a rough estimate, valid up to a
numerical coefficient, of the actual scale of non--perturbative
physics. An explicit calculation of the scattering amplitude,
including inelastic channels, shows that this is indeed the case
and the numerical factor is found to be roughly
$1/4$~\cite{Papucci}.

Since the ratio of the $W$ to the first KK mode mass squared is of
order \beq \frac{M_W^2 }{M_{W^{(1)}}^2}= \mathcal{O} \left(
{1}/{\log  \left({R^\prime}/{R}\right)} \right)~, \eeq raising the
value of $R$ (corresponding to lowering the 5D UV scale) will
significantly increase the NDA cutoff. With $R$ chosen to be the
inverse Planck scale, the first KK resonance appears around
$1.2$~TeV, but for larger values of $R$ this scale can be safely
reduced down below a TeV.

\section{Fermions in extra dimensions}
\label{sec:ferm} \setcounter{equation}{0} \setcounter{footnote}{0}

Since the early 1980's, there was a well known issue with allowing
fermions to propagate in the bulk of an extra dimension. This
problem arises from the spin-1/2 representations of the Lorentz
group in higher dimensions.  The principle issue is that the
irreducible representations of the Lorentz group in higher
dimensional spaces are not necessarily chiral from the 4D point of
view.  This means that a low energy effective theory derived from
this higher dimensional theory would not, in general, contain
chiral fermions.  However, the SM does contain chiral fermions,
and so models with bulk fermions appeared to be doomed.  It was
realized in the 80's, however, that it is possible to obtain
chiral fermions in models with extra dimensions via orbifolding.
Our first objective will be to outline how this is possible.  We
cover cases where the background geometry of the extra dimension
is either flat or
warped~\cite{neubert,GherPom,HS1,nomurasmith,HSrecent}, and give
some explicit examples of interesting models with bulk
fermions~\cite{kaptait}.  The standard orbifold method of
producing chiral modes is generalized to include arbitrary
fermionic boundary conditions.  In Higgsless models, the boundary
condition techinique is used  to generate the entire spectrum of
SM fermion masses through boundary conditions~\cite{CGHST}. We
also discuss a simpler model of obtaining fermions masses through
localization methods.  It is interesting to note that extra
dimensions have provided an alternative framework to possibly
resolve the flavor hierarchy problem of the standard model.

\subsection{Brief Summary of Fermions in D$=4$}

Before we begin our discussions of fermions in higher dimensions,
we first review the basic properties of 4D fermions~\cite{theHungarians}.  We follow the
spinor conventions given in~\cite{WB}.  We use the
chiral representation for the Dirac $\gamma$ matrices:
\begin{equation}
\gamma^\mu = \left(\begin{array}{cc}
0 & \sigma^\mu \\
\bar{\sigma}^\mu & 0
\end{array}\right)\ \  \mathrm{and}\ \
\gamma^5 = \left(\begin{array}{cc}
i & 0 \\
0 & -i
\end{array}\right)\ \ \mu = 0,1,2,3~,
\end{equation}
where $\sigma^i=-\bar{\sigma}^i$ are the usual Pauli spin
matrices, while $\sigma^0 = \bar{\sigma}^0 = -{\mathbf 1}$.

The Lorentz group of 4-vectors $x^\mu$ is defined through the
transformation
\begin{equation}
x'^\mu = \Lambda^\mu_\nu x^\nu
\end{equation}
where the $\Lambda$ matrices are such that they leave the
Minkowski inner products invariant:
\begin{equation}
x'^\mu  g'_{\mu \nu} y'^\nu = x^\mu g_{\mu \nu} y^\nu
\end{equation}

Spinors are a different type of representation of the Lorentz
group. To start of the discussion of spin-1/2 representations, we
note that in 4 dimensions, the 2-dimensional complex special
linear group, $SL(2, \mathbb{C})$, can be shown to be a covering
space for the Lorentz group.  This equivalence is similar to the
mapping of the special unitary group, $SU(2)$, onto the rotation
group $SO(3)$. To see this more explicitly, consider the following
parametrization of a Lorentz four-vector $x^\mu$
\begin{equation}
x^\mu \rightarrow [ x ] \equiv x^0 - x^i \sigma^i =
\left[\begin{array}{cc}
x^0 - x^3 & - x^1 + i x^2 \\
-x^1-ix^2 & x^0 + x^3
\end{array}\right],
\end{equation}
where $[x]$ has the following properties:
\begin{equation}
[x] = [x]^\dagger\ \ \mathrm{and}\ \ \det [x] = x^\mu x^\nu
g_{\mu\nu} \equiv x^\mu x_\mu.
\end{equation}

Now take an arbitrary matrix $A \in SL(2,\mathbb{C})$.  Such a
matrix is a general $2 \times 2$ complex matrix with unit
determinant.  Under a rotation by $A$,
\begin{equation}
[x] \rightarrow [x]_A = A [x] A^\dagger
\end{equation}
Finally note that $[x]_A = [x]_A^\dagger$, and that $\det [x]_A =
\det [x]$.

These all correspond precisely to the properties of the inner
product under a general Lorentz transformation.  Thus, for some
$\Lambda_A$,
\begin{equation}
[x]_A = [\Lambda_A x].
\end{equation}
Thus the mapping $A \rightarrow \Lambda_A$ is a homomorphism of
the Lorentz group.

The group $SL(2,\mathbb{C})$ is isomorphic to the product group
$SU(2) \times SU(2)$, with the transformation parameters for the
$SU(2)$'s being complex, but related by complex conjugation.  The
imaginary component of the transformation parameters is associated
with the non-compact directions of the Lorentz group (the boosts)
while the rotations are associated with the real part of these
parameters. Because of this isomorphism, we can express
representations of $SL(2,\mathbb{C})$ in terms of their breakdown
under the $SU(2)$ subgroups.  The two $SU(2)$ indices are
represented as dotted and un-dotted.  The two most simple
(non-trivial) irreducible representations of $SL(2,\mathbb{C})$
can then be written as $\chi^\alpha$ and
$\bar{\psi}^{\dot{\alpha}}$. To introduce a notation, these are
the $(1/2,0)$ and $(0,1/2)$ representations of the Lorentz group,
respectively.  These are the familiar left and right handed Weyl
spinors.

Any representation of the Lorentz group can then be written in
terms of the transformation laws under the two complexified
$SU(2)$ subgroups. Such a representation is labelled $(m,n)$ where
$m$ is the number of un-dotted indices that the representation
has, while $n$ is the number of dotted indices.

As a side note, we mention that if we require that a physical
theory be invariant under a parity transformation, then we require
that it contain special combinations of fundamental
representations.  Parity exchanges left and right handedness, or
in terms of the labelling $(m,n)$ of a representation, exchanges
$m$, and $n$.  For a theory to be invariant under parity, it must
contain representations in the form of direct sums $(m,n)
\Circleplus (n,m)$.  In the case of Weyl spinors, the lowest
representation that is invariant under parity is the $(1/2,0)
\Circleplus (0,1/2)$.  This representation is the familiar Dirac
spinor.

In the index notation that we have discussed, the rules for the
indices are quite simple.  Complex conjugation, exchanges dotted
and undotted indices.  The metrics on the dotted and un-dotted
spaces which raise and lower indices are given by the
anti-symmetric tensors $\epsilon_{\alpha \beta}$ and
$\epsilon_{\dot{\alpha} \dot{\beta}}$.  In terms of the
$SL(2,\mathbb{C})$ notation, the $\sigma$ matrices exchange
representations between the dotted and undotted spaces:
\begin{equation}
\sigma^\mu \rightarrow \sigma^\mu_{ \alpha \dot{\alpha}}
\end{equation}
This follows from the transformation properties of $[x]$.  With
these conventions, spinors can be combined into invariants which
have the following property:
\begin{equation}
\chi^\alpha \psi^\beta \epsilon_{\alpha \beta} \equiv \chi^\alpha
\psi_\alpha = \psi^\alpha \chi_\alpha.
\end{equation}
There are two minus signs that cancel each other in switching the
ordering of the two spinors.  One is from changing the order of
the Grassman variables making up the spinors, and the other is
from permuting the indices in the totally anti-symmetric tensor,
$\epsilon_{\alpha \beta}$. Because the spinor sums can be
interchanged in this way, in the proceeding sections we will
frequently drop the spinor indices completely:  $\chi^\alpha
\psi_\alpha \equiv \chi \psi$.  Note, however, that $\chi^\alpha
\psi_\alpha = - \chi_\alpha \psi^\alpha$.
\subsection{Fermions in a Flat Extra Dimension}
\label{sec:BCs} \setcounter{equation}{0}
In 5D, the Clifford algebra includes, in addition to the four
dimensional Dirac algebra, a $\gamma^5$.  This $\gamma^5$ is
precisely the parity transformation that we discussed in the
previous section.  This means that in 5D the simplest irreducible
represention will break up under the 4D subgroup of the full 5D
Poincar\'e algebra as a $(0,1/2) \Circleplus (1/2,0)$.  That is,
the simplest irreducible representation in 5D is a Dirac spinor,
rather than a Weyl spinor.  This is expressing the fact that bulk
fermions are not chiral, as mentioned in the introduction to this
lecture.  It is not possible to start only with 2 component
spinors, as can be done in 4D theories.

As a warmup to working in more general compactified spaces, we
consider the minimal 5D Lagrangian for a bulk spinor field which
is propagating in a flat extra dimension with the topology of an
interval::
\begin{equation}
\label{eq:BulkAction} S = \int d^5 x \left( \frac{i}{2} (
\bar{\Psi}\, \Gamma^M \partial_M \Psi - \partial_M \bar{\Psi}\,
\Gamma^M  \Psi ) - m \bar{\Psi} \Psi  \right).
\end{equation}
The field $\Psi$ decomposes under the 4D Lorentz subgroup into two
Weyl spinors:
\begin{equation}
\Psi = \left(
\begin{array}{c} \chi_\alpha
\\
\tv{13} \bar{\psi}^{\dot{\alpha}}
\end{array}
\right). \label{bigpsi}
\end{equation}
In finding the consistent boundary conditions for these Weyl
fermions, it is useful to express the Lagrangian
(\ref{eq:BulkAction}) in terms of the 4D Weyl spinors:
\begin{equation}
S = \int d^5 x \left( - i \bar{\chi} \bar{\sigma}^{\mu}
\partial_\mu \chi - i \psi \sigma^{\mu} \partial_\mu \bar{\psi}
+\sfrac{1}{2}\, ( \psi  \overleftrightarrow{\partial_5}  \chi -
\bar{\chi}  \overleftrightarrow{\partial_5}   \bar{\psi}   ) + m
(\psi \chi +  \bar{\chi} \bar{\psi} ) \right),
\end{equation}
where $\overleftrightarrow{\partial_5} =
\overrightarrow{\partial_5}-\overleftarrow{\partial_5}$.

We note that in 4D theories, the terms with the left acting
derivatives are generally integrated by parts, so that all
derivatives act to the right.  However, since we are working here
in a compact space with boundaries, the integration by parts
produces boundary terms which can not be neglected.

The bulk equations of motion for the 4D Weyl spinors which result
from the variation of this 5D Lagrangian are:
\begin{eqnarray}
\label{bulkeom} -i \bar{\sigma}^{\mu} \partial_\mu \chi -
\partial_5 \bar{\psi} + m \bar{\psi} = 0,
\nonumber \\
-i \sigma^{\mu} \partial_\mu \bar{\psi} + \partial_5 \chi + m \chi
= 0.
\end{eqnarray}
\subsection{Boundary Conditions for Fermions in 5D}

Our goal now is to find what the possible consistent boundary
conditions are.  We consider a consistent boundary condition to be
one which satisfies the action principle.  Naively, one might
think that there are two independent spinors, $\chi$ and
$\bar{\psi}$, and that one would require two independent boundary
conditions for each spinor.  However, because the bulk equations
of motion are only first order, there is only one integration
constant.  So for the Dirac pair, $(\chi, \bar{\psi})$, there is
only one boundary condition $f(\chi,\psi) = 0$ at each boundary,
where $f$ is some function of the spinors and their conjugates.
The form of $f$ together with the bulk equations of motion in
Eq.~(\ref{bulkeom}) then determines all of the arbitrary
coefficients in the complete solution to the spinor equation of
motion on the interval.

We would now like to see what the restrictions are on the function
$f$, so now let us examine the variations that include the
derivatives acting along the extra dimension:
\begin{equation}
\delta S = \int \frac{1}{2} \left( \delta \psi
\overleftrightarrow{\pd_5} \chi + \psi \overleftrightarrow{\pd_5}
\delta \chi - \delta \bar{\chi} \overleftrightarrow{\pd_5}
\bar{\psi} - \bar{\chi} \overleftrightarrow{\pd_5} \delta
\bar{\psi} \right)
\end{equation}
To get the boundary equations of motion, we need to integrate by
parts so that there are no derivatives acting on the variations of
the fields left over. However, this procedure results in residual
boundary terms given by
\begin{equation}
\delta S^{\mathrm{bound}} = \int d^5 x \left[ - \delta \psi \chi +
  \psi \delta \chi + \delta \bar{\chi} \bar{\psi} - \bar{\chi} \delta
  \bar{\psi} \right]^L_0
\end{equation}

The most general boundary conditions which satisfy the action
principle then are given by the solutions to
\begin{equation}\label{bcbasic}
- \delta \psi \chi +
  \psi \delta \chi + \delta \bar{\chi} \bar{\psi} - \bar{\chi} \delta
  \bar{\psi} = 0.
\end{equation}
As a simple example, consider the case when the spinors are
proportional to each other on the boundaries:
\begin{equation}
\psi = \alpha \chi.
\end{equation}
The variations of the spinors are then related by
\begin{equation}
\delta \psi = \alpha \delta \chi
\end{equation}
Plugging these relations into (\ref{bcbasic}), we find that it
simplifies to
\begin{equation}
\alpha \delta \chi \chi - \alpha \delta \chi \chi = 0,
\end{equation}
and the variation of the action on the boundaries does indeed
vanish.

This is only one of many boundary conditions which satisfy the
action principle.  The most general solution consistent with the
Lorentz symmetries of the interval is given by
\begin{equation}
\psi_\alpha = M_\alpha^\beta \chi_\beta + N_{\alpha \dot{\beta}}
\bar{\chi}^{\dot{\beta}}.
\end{equation}
We have put the Weyl indices back in to show the structure of the
operators $M$ and $N$.  These operators can contain derivatives
along the extra dimension.

This most general set of boundary conditions is further restricted
by additional symmetries such as gauge symmetries that are allowed
on the boundaries.  For example, if a fermion is transforming
under a complex representation of a gauge group, then the operator
$M_\alpha^\beta$ must vanish.  This is because the spinors $\chi$
and $\psi$ transform under conjugate representations, thus the
spinors cannot consistently be proportional to each other.  If the
fields are in real representations of the gauge group, such as the
adjoint, such boundary conditions are allowed.

Let us consider a simple boundary condition: take the spinor
$\psi$, and set it equal to zero on both boundaries.  The
resulting boundary condition for the other Weyl spinor $\chi$,
which comes from the bulk equation of motion, is
\begin{equation}
\left( \pd_5 + m \right) \chi |_{0,L} = 0.
\end{equation}
Solving the equations of motion with these boundary conditions
results in a zero mode for $\chi$, but not for $\psi$.  That is,
the low energy theory is a chiral theory, which has been obtained
from an inherently non-chiral 5D theory.  By appropriately
choosing the boundary conditions, one can get a chiral effective
theory from a geometry which would naively not allow chiral modes.

\begin{figure}[t]
\centerline{\includegraphics[width=0.75\hsize]{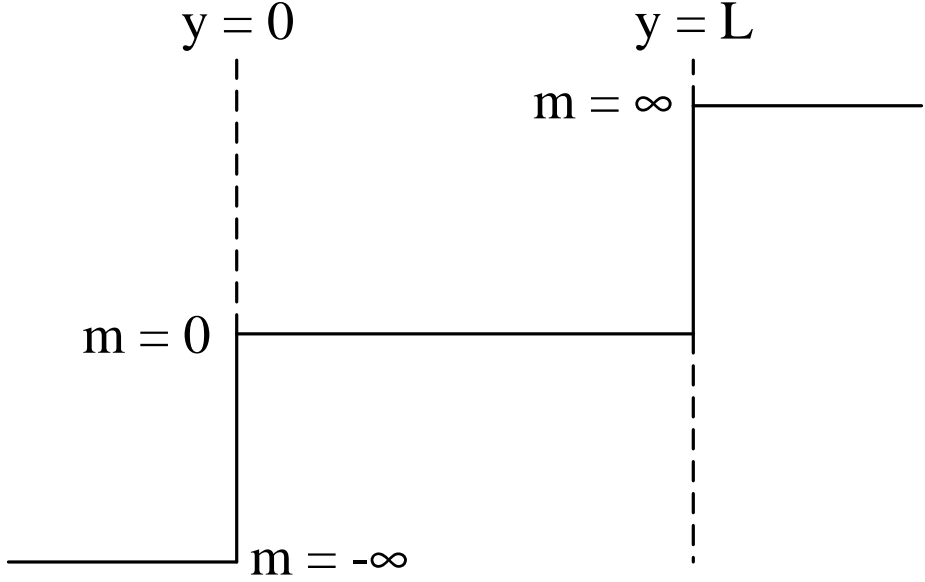}}
\caption{A fermion mass profile in an infinite extra dimension
which
  leads to a 4 dimensional chiral zero mode.  The resulting zero mode
  wave function is
  flat and finite in the interval $0 < y < L$.  The wave function for
  this zero mode is vanishing at all other points in the extra dimension.}
\label{fig:phys}
\end{figure}

It is useful to have in mind a physical picture which could result
in this type of boundary condition.  For this purpose, we can
consider an infinite extra dimension where there is a finite
interval where the bulk Dirac mass is vanishing, but outside of
which the mass is either positive and infinite, or negative and
infinite.  Then a constant mass $m$ is added.  This is shown
pictorially in Figure \ref{fig:phys}.  In the case where the sign
of the mass is opposite on either end of the interval, after
solving the bulk equations in the entire bulk space, we get the
boundary condition above at the points $y=0$ and $y=L$ that
resulted in a zero mode for the Weyl spinor, $\chi$.  This mass
profile is a discretized version of the boundary wall localized
chiral fermion approach~\cite{Kaplan,arkschmaltz}.

In the case where the Dirac mass is the same sign on either end of
the interval $0<y<L$, the fermions are again localized, however
the boundary conditions are $\chi |_{y=0} = 0$ and $\psi |_{y=L} =
0$.  In this case, no zero mode results, and the lowest lying KK
resonance has a mass of order $1/L$.

To show that the orbifold picture does not easily give all
possibilities, let us consider this same example in the orbifold
language.  In the orbifold setup, the boundary conditions are
determined by imposing $\Z_2$ parity ($y \rightarrow -y$) symmetry
on the spinors, where the $\psi$ spinor is odd, and the $\chi$
spinor is even.  The parity transformation of one spinor is then
determined by the other, since the action contains terms of the
form $\psi \pd_5 \chi$.  If $\psi$ is odd under the $\Z_2$ then
$\chi$ must be even, since $\pd_5 \rightarrow - \pd_5$ under the
parity transform.  The situation becomes complicated, however, if
one wants to give a bulk Dirac mass to the 5D fermion: $M \psi
\chi$.  This term is not allowed under the $Z_2$ symmetry unless
the bulk mass term is given a transformation law under the $Z_2$
as well.  This means that the bulk mass term must undergo a
discrete jump at the orbifold fixed points~\cite{BFZ,otherBagger}
(those points that are stationary under the parity
transformation).  In the interval picture, there are no such
issues.  The vanishing of the boundary and bulk action variation
give the requirements that $\psi = 0$, and $(\pd_5 + m) \chi = 0$
at the endpoints.

\subsection{Examples and a Simple Application}
We begin this section with a discussion of the KK-decomposition of
the 5D fermion fields, which we then apply to an application which
provides an interesting solution to the fermion mass hierarchy
problem, the Kaplan-Tait model~\cite{kaptait}.  This approach utilized
the boundary wall fermion localization method~\cite{Kaplan}.

As with gauge and scalar fields, there will be, in the 4D
effective theory, a tower of massive Dirac fields that arise from
solving the full 5D spectrum.  These fermions will obey the 4D
Dirac equation, which, when broken into Weyl spinors, is given by:
\begin{eqnarray}
- i \bar{\sigma}^\mu \pd_\mu \chi^{(n)} + m_n \bar{\psi}^{(n)} = 0
  \arline
- i \sigma^\mu \pd_\mu \bar{\psi}^{(n)} + m_n \chi^{(n)} = 0
\end{eqnarray}
The 5D spinors $\chi$ and $\psi$ can be written as a sum of
products of the 4D Dirac fermions with 5D wavefunctions:
\begin{eqnarray}
        \label{eq:DiracKK}
\chi  = \sum_n g_n(y)\, \chi_{n} (x), \\
\bar{\psi} = \sum_n f_n(y)\, \bar{\psi}_n (x).
\end{eqnarray}
Substituting this decomposition into the 5D bulk equations of
motions gives the following

\begin{eqnarray}
g_n' + m\, g_n - m_n\, f_n = 0, \\
f_n' - m\, f_n + m_n\, g_n = 0. \label{eq:1storder}
\end{eqnarray}

The standard approach to solving this system of equations is to
combine the two first order equations into two second order wave
equations:
\begin{eqnarray}
g''_n +(m_n^2-m^2) g_n = 0,\\
f''_n +(m_n^2-m^2) f_n = 0.
\end{eqnarray}
The solution is simply a sum of sines and cosines, with
coefficients that are determined by reimposing the first order
equations, and imposing the boundary conditions.

In the Kaplan-Tait model, there is a Higgs field which is confined
to one boundary of an extra dimension, and there are gauge fields
which are propagating in the bulk.  Assign the bulk fermions the
boundary conditions where $\psi |_{0,L} = 0$, and $(\pd_5 + m)
\chi |_{0,L} = 0$.  The main question concerns the zero mode
solutions.  Take the first order equations (\ref{eq:1storder}),
and set the 4D mass eigenvalue to zero.  The resulting equations
are
\begin{eqnarray}
g_n' +m g_n = 0 \arline f_n' - m f_n = 0.
\end{eqnarray}
The solutions are simply exponentials.  The solution which obeys
the boundary condition $\psi |_{0,L} = 0$ is $f_0 = 0$ and $g_0
(y) = g_0 e^{-m y}$.  The wave-function then either exponentially
grows or decays, depending on whether the bulk mass term is
positive or negative.  The constant $g_0$ is determined by the
choice of normalization for the fermion wave function.  To obtain
a 4D theory in which the zero mode has the canonical
normalization, we impose that
\begin{equation}
\int_0^L g_0^2 (y) dy = 1,
\end{equation}
which has the solution
\begin{equation}
g_0 = \sqrt{\frac{2 m}{1- e^{-2 m L}}}.
\end{equation}

\begin{figure}[t]
\centerline{\includegraphics[width=0.5\hsize]{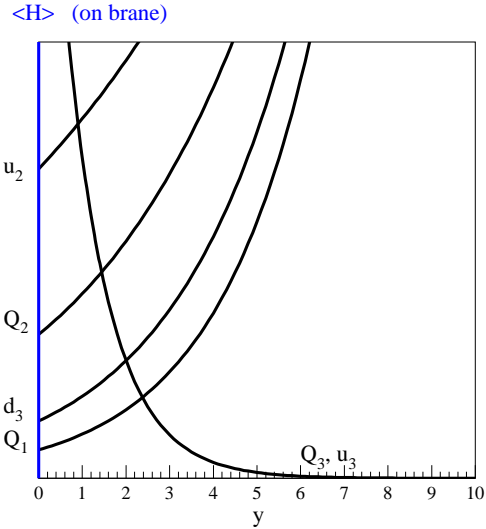}}
\caption{This figure, from~\cite{kaptait}, displays how the
  fermion mass hierarchy is achieved through localization of the
  chiral zero modes.  The lighter quarks are localized away from
  $y=0$, while the third generation is peaked on the $y=0$ brane, so
  that it has a sizable overlap with the Higgs VEV.}
\label{fig:kaptait}
\end{figure}

Now we can propose that all of the Yukawa couplings of the bulk
fermions to the Higgs on the boundary are of order one, and try to
find what the masses are for different bulk Dirac masses.  The
Yukawa couplings in the 5D theory are given by
\begin{equation}
\lambda_u L \bar{u}_R H Q_L \delta (y-L) \rightarrow \lambda_u L
\bar{\chi}_Q^0 H \chi^{0*}_u \delta (y-L)
\end{equation}
where the expression on the right leaves out all modes except the
zero mode left from the solution above.  The effective Yukawa
coupling in the 4D picture involves the wave function evaluated at
the boundary where the Higgs is located, and is expressed as (for
example):
\begin{equation}
\lambda_{4D}^u = \frac{\lambda_{5D}^u}{\sqrt{2}} \frac{ \sqrt{m_Q
m_u
    L^2} }{ \sqrt{ \left( 1- e^{-2 m_Q L} \right) \left( 1- e^{-2 m_u L}
    \right) } } e^{-(m_Q + m_u) L}.
\end{equation}
This turns out to be an interesting solution to the fermion mass
hierarchy. For all parameters of order one, it is possible to get
a very wide spectrum of fermion masses.  This is due to the
exponential dependence of the wave functions on the bulk masses.
For a small range of bulk Dirac masses that are all ${\mathcal O}
(1)$, one can easily lift the zero modes. The exponential
dependance of the effective 4D Yukawa coupling on the bulk Dirac
mass implies that this small range can give the fermions 4D masses
that span the observed standard model spectrum.  A graphical
representation of how this works is given in
Figure~\ref{fig:kaptait}.

\subsubsection{Fermions in Warped Space}

As shown in earlier parts of these lectures, the tools of warped
spacetime could fix some of the phenomenological problems of flat
extra dimensions, and we would now like to see whether we can
replicate the standard model spectrum of fermions in AdS setups.
The first complication is that we need to know the form of the
covariant derivative acting on fermions in this curved spacetime.

To this end, we need an object constructed from the metric which
lives in the spin 1/2 representation of the Lorentz group.  In
very rough terms, this object is the square root of the metric,
$g$.  In index notation, we write down the metric in terms of
``vielbeins,'' or in the case of 5 dimensions, a ``f\"unfbein'':
\begin{equation}
g^{MN} = e_a^M \eta^{ab} e_b^N
\end{equation}
The Dirac algebra is written in curved space in terms of the flat
space Dirac matrices as
\begin{equation}
\Gamma^M = e^M_a \gamma^a
\end{equation}

To write down the covariant derivative that can act on fermions,
we use a spin connection, $w_M^{ab}$:
\begin{equation}
D_M = \pd_M + \frac{1}{2} w_M^{ab} \sigma_{ab},
\end{equation}
where $\sigma_{ab} = \frac{1}{4} \gamma_{\left[ a \right.}
  \gamma_{\left. b \right]}$.
The spin connection can be expressed in terms of the f\"unfbeins
as~\cite{spinconn}
\begin{equation}
w_M^{ab} = \frac{1}{2} g^{RP} e^{\left[ a \right.}_R \pd_{\left[ M
      \right.} e^{\left. b \right]}_{\left. p \right]}
+ \frac{1}{4} g^{RP} g^{TS} e^{\left[ a \right.}_R e^{\left. b
      \right]}_T \pd_{\left[ S \right.} e^c_{\left. P \right]} e^d_M \eta_{cd}.
\end{equation}

When the background geometry is given by AdS space, the metric (in
conformal coordinates) is given by
\begin{equation}
ds^2 = \left( \frac{R}{z} \right)^2 \left(dx_\mu dx_\nu \eta^{\mu
\nu} - dz^2 \right).
\end{equation}
One can show that $e^a_M = \left( \frac{R}{z} \right) \delta^a_M$,
$D_\mu \psi = \left( \pd_\mu + \gamma_\mu \gamma_5 \frac{1}{4z}
\right) \psi$, and $D_5 \psi = \pd_5 \psi$. Proving that the spin
connection terms cancel each other in this manner is left as an
exercise for the reader.

Written in terms of the two component Weyl spinors, the AdS action
will be
\begin{equation}
S = \int d^5 x \left(\frac{R}{z}\right)^4
 \left(
- i \bar{\chi}  \bar{\sigma}^\mu \partial_\mu \chi - i \psi
\sigma^{\mu} \partial_\mu \bar{\psi} + \sfrac{1}{2} ( \psi
\overleftrightarrow{\partial_5} \chi -  \bar{\chi}
\overleftrightarrow{\pd_5} \bar{\psi} ) + \frac{c}{z} \left( \psi
\chi + \bar{\chi} \bar{\psi} \right) \right),
\end{equation}
where the coefficients $c = m R$, and $m$ is the bulk Dirac mass
term for the 4-component Dirac spinor.  In AdS space, the bulk
equations of motion are~\cite{neubert}:
\begin{eqnarray}
\label{bulkeq1} -i \bar{\sigma}^{\mu} \pd_\mu \chi - \pd_5
\bar{\psi} + \frac{c+2}{z} \bar{\psi} = 0,
 \\
-i \sigma^{\mu} \pd_\mu \bar{\psi} + \pd_5 \chi + \frac{c-2}{z}
\chi = 0. \label{bulkeq2}
 \end{eqnarray}
These have some subtle yet important features. The terms in the
equations of motion that contain the bulk mass, $c$, are dependent
on the extra dimensional coordinate, $z$.  The $z$ dependent terms
$2/z$ play an important role in determining the localization of
any potential zero modes.

As before, we perform the KK decomposition.  Everything from flat
space carries over, except that the bulk equations of motion for
the wave functions are different.
\begin{equation}
\chi = \sum_n g_n (z) \chi_n (x) \ \ \mathrm{and} \ \ \bar{\psi} =
\sum_n f_n (z) \bar{\psi}_{n} (x),
\end{equation}
where the 4D spinors $\chi_n $ and $\bar{\psi}_{n}$ satisfy the
usual 4D Dirac equation with mass $m_n$:
\begin{equation}
-i \bar{\sigma}^\mu \partial_\mu \chi_n + m_n \bar{\psi}_n = 0 \ \
\mathrm{and} \ \ -i \sigma^\mu \partial_\mu \bar{\psi}_n + m_n
\chi_n = 0.
\end{equation}
The bulk equations then become ordinary (coupled) differential
equations of first order for the wavefunctions $f_n$ and $g_n$:
\begin{eqnarray}
        \label{eq:beom1}
& \displaystyle f^\prime_n + m_n g_n - \frac{c+2}{z} f_n = 0,
\\
        \label{eq:beom2}
& \displaystyle g^\prime_n  - m_n g_n + \frac{c-2}{z} g_n = 0.
\end{eqnarray}

For a zero mode, if the boundary conditions were to allow its
presence, these bulk equations are already decoupled and are thus
easy to solve, leading to:
\begin{eqnarray}
& \displaystyle f_0  = C_0 \left( \frac{z}{R} \right)^{c+2},
\\
& \displaystyle g_0 = A_0  \left( \frac{z}{R} \right)^{2-c},
\end{eqnarray}
where $A_0$ and $C_0$ are two normalization constants of mass
dimension $1/2$.

For the massive modes, just as with the flat space scenario, the
first order differential equations can be uncoupled by combining
them to get second order equations:
\begin{eqnarray}
& \displaystyle f^{\prime\prime}_n - \sfrac{4}{z} f^\prime_n +
(m_n^2 - \sfrac{c^2-c-6}{z^2}) f_n = 0,
\\
& \displaystyle g^{\prime\prime}_n - \sfrac{4}{z} g^\prime_n +
(m_n^2 - \sfrac{c^2+c-6}{z^2}) g_n = 0.
\end{eqnarray}
The solutions are now linear combinations of Bessel functions, as
opposed to $\sin$ and $\cos$ functions:
\begin{eqnarray}
        \label{eq:Bessel1}
& \displaystyle g_n(z) = z^\frac{5}{2} \left( A_n
J_{c+\frac{1}{2}}(m_n z) + B_n Y_{c+\frac{1}{2}}(m_n z) \right)
\\
        \label{eq:Bessel2}
& \displaystyle f_n(z) = z^\frac{5}{2} \left( C_n
J_{c-\frac{1}{2}}(m_n z) + D_n Y_{c-\frac{1}{2}}(m_n z) \right).
\end{eqnarray}
The first order bulk equations of motion
(\ref{eq:beom1})-(\ref{eq:beom2}) further impose that
\begin{equation}
A_n = C_n \ \ \mathrm{and} \ \ B_n = D_n.
\end{equation}
The remaining undetermined coefficients are determined by the
boundary conditions, and the wave function normalization.

As in the flat space Kaplan-Tait model, the bulk mass determines
the localization of the fermions.  For instance, let us take
\begin{equation}
g_0 (y)  = A_0 \left( \frac{z}{R} \right)^{2-c}, \ \ f=0.
\end{equation}
This solution corresponds to the boundary condition where $\psi
|_{R,R'} = 0$.  The coefficient $A_0$ is determined by the
normalization condition
\begin{equation}
\int_R^{R'} dz \left( \frac{R}{z} \right)^5 \frac{z}{R} A_0^2
\left( \frac{z}{R} \right)^{4-2c} = A_0^2 \int^{R'}_R \left(
\frac{z}{R} \right)^{-2c} dz=1.
\end{equation}
To understand from these equations where the fermions are
localized, we study the behavior of this integral as we vary the
limits of integration.  For the boundary conditions that we are
studying right now, if we send $R'$ to infinity, we see that the
integral remains convergent if $c > 1/2$, and the fermion is then
localized on the UV brane.  If we send $R$ to zero, the integral
is convergent if $c < 1/2$, and the fermion is localized on the IR
brane.  The value of the Dirac mass determines whether the fermion
is localized towards the UV or IR branes.  We note that the
opposite choice of boundary conditions that yields a zero mode
($\chi |_{R, R'} = 0$) results in a zero mode solution for $\psi$
with localization at the UV brane when $c< -1/2$, and at the IR
brane when $c>-1/2$. The interesting feature in the warped case is
that the localization transition occurs not when the bulk mass
passes through zero, but at
 points where $|c| = 1/2$.  This is due to the curvature effects of
 the extra dimension.

\subsection{Fermion Masses in the Higgsless Model}

Recall the gauge symmetries of the Higgsless model in warped space
shown in figure \ref{fig:higgsless}.  The fermions in this model
can not be completely localized on the UV or IR branes.  If they
were on the IR brane, the up-type and down-type quarks could not
have any mass splitting as the theory is non-chiral at $z=R'$, and
if they were on the UV brane, the theory would be completely
chiral, and the zero modes could not be lifted.  The fermions must
then live in the bulk, and feel the breakings of both branes.  The
quantum numbers of the fermions are given in
Table~\ref{tab:quantnums}.
\begin{table}
\center{\begin{tabular}{cccc}
$$      &   $SU(2)_L$   &    $SU(2)_R$    &   $U(1)_{B-L}$ \arline
$\left(\begin{array}{c}u \\d \end{array}\right)_L$   &   $\Box$
&    $1$          &   $1/6$        \arline
$\left(\begin{array}{c}u \\ d \end{array}\right)_R$
        &   $1$         &    $\Box$       &   $1/6$        \arline
$\left(\begin{array}{c}\nu \\ e \end{array}\right)_L$
        &   $\Box$      &    $1$          &   $-1/2$       \arline
$\left(\begin{array}{c}\nu \\ e \end{array}\right)_R$
        &   $1$         &    $\Box$       &   $-1/2$
\end{tabular}}
\caption{These are the quantum numbers of the bulk fermions under
the
  bulk left-right symmetric Higgsless model.}\label{tab:quantnums}
\end{table}
The preliminary boundary conditions for the fermions that give the
zero modes that we desire are given by
\begin{equation}
\left(\begin{array}{c} \chi_{u_L} \\ \bar{\psi}_{u_L} \\ \chi_{d_L} \\
\bar{\psi}_{d_L} \end{array}\right)
\begin{array}{cc} + & + \\ -&- \\ +&+ \\ -&- \end{array}
\ \ \ \ \ \ \
\left(\begin{array}{c} \chi_{u_R} \\ \bar{\psi}_{u_R} \\ \chi_{d_R} \\
\bar{\psi}_{d_R} \end{array}\right)
\begin{array}{cc} -&- \\ +&+ \\ -&- \\ +&+ \end{array}.
\label{eq:zeromodes}
\end{equation}
Where the $+$ and $-$ refer to whether we give those spinors
Neumann or Dirichlet boundary conditions, respectively.  These
boundary conditions give massless chiral modes that match the
fermion content of the standard model. However, the $u_L$, $d_L$,
$u_R$, and $d_R$ are all massless at this stage, and we need to
lift the zero modes to achieve the standard model mass spectrum.
While simply giving certain boundary conditions for the fermions
will enable us to lift these zero modes, in the following
discussion, we talk about boundary operators, and the boundary
conditions that these operators induce.  There are some subtleties
in dealing with boundary operators for fermions.  These arise from
the fact that the fields themselves are not always continuous in
the presence of a boundary operator.  This is due to the fact that
the equations of motion for fermions are first order.  The most
straightforward approach is to enforce the boundary conditions
that give the zero modes as shown in Eq. (\ref{eq:zeromodes}) on
the real boundary at $z=R,R'$ while the boundary operators are added
on a fictitious brane a distance $\epsilon$ away from it.  The distance between the fictitious brane and the
physical one is taken to be $\epsilon$. The new boundary condition
is then obtained by taking the distance $\epsilon$ to be small.
This physical picture is quite helpful in understanding what the
different boundary conditions will do.

To lift a doublet, we can give Dirac masses on the TeV brane which
mix the $SU(2)_L$ and $SU(2)_R$ multiplets.  This is possible
because the theory on the IR brane is non-chiral.  The boundary
conditions that this Dirac mass gives on the IR brane are
\begin{equation}
\label{eq:dmassbc} \psi_L = - M_D R' \psi_R \ \ \ \mathrm{and} \ \
\ \chi_R = M_D R' \chi_L
\end{equation}

At this stage, the up and down-type quarks, and the charged
leptons and neutrinos are degenerate in mass.  Mass splittings
must be acquired on the UV brane, where the theory is chiral.  For
leptons and neutrinos, this can be accomplished simply by adding a
Majorana mass for the neutrinos on the UV brane.  The neutrino
mass is suppressed by a type of see-saw mechanism while the lepton
masses are unaffected.

\begin{figure}[t]
\centerline{\includegraphics[width=0.75\hsize]{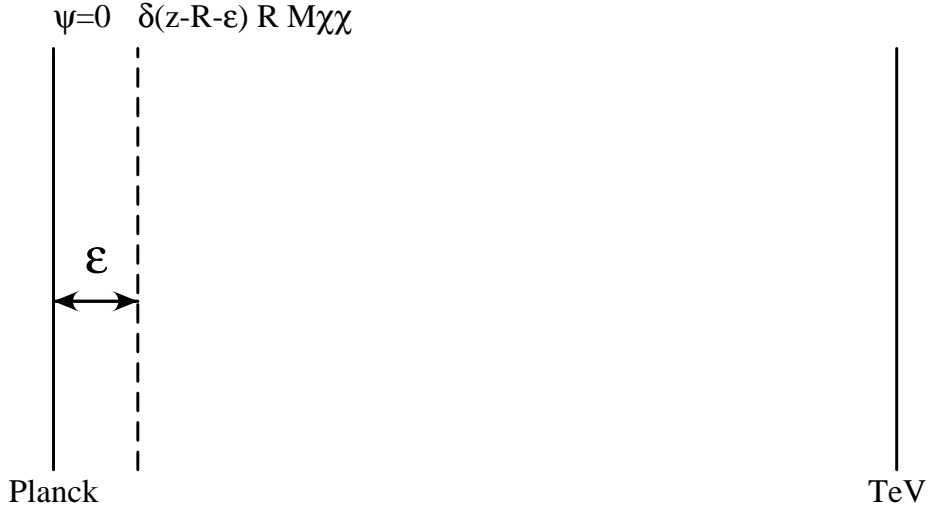}}
\caption{To give a physical picture to the boundary conditions
used to lift a zero mode, a 4D majorana mass is added a slight
distance from the boundary.  The equations of motion are then
solved, and the limit $\varepsilon \rightarrow 0$ is taken.}
\label{fig:majmass}
\end{figure}

As a complete example of the boundary operator prescription, we
consider adding a Majorana mass for the $\nu_R$ on the UV brane.
This is displayed in Figure \ref{fig:majmass}.  Going through the
procedure described above, the new equation of motion is given by
\begin{equation}
-i \bar{\sigma} \pd_\mu \chi - \pd_5 \bar{\psi} + \frac{c+2}{z}
 \bar{\psi} + \frac{M R^2}{z} \bar{\chi} \delta (z-R-\epsilon) = 0
\end{equation}
Integrating over the discontinuity, we obtain a jump condition for
the spinor $\psi$:
\begin{equation}
\left[ \psi \right]^{R+\epsilon}_{R} = M R \chi |_{R+\epsilon}.
\end{equation}
With the condition that $\psi |_R = 0$ and the small $\epsilon$
limit, we have a new effective boundary condition
\begin{equation}
\label{eq:majmassbc} \psi_{R+\epsilon} = M R \chi |_{R+\epsilon}
\end{equation}

The general solutions to the bulk equations of motion are given
by:
\begin{eqnarray}
\chi_{L,R} = z^{5/2} \left[ A_{L,R} J_{1/2+c_{L,R}} (m z) +
B_{L,R}
    J_{-1/2 - c_{L,R}} (m z) \right] \arline
\psi_{L,R} = z^{5/2} \left[ A_{L,R} J_{-1/2+c_{L,R}} (m z) -
B_{L,R}
    J_{1/2 - c_{L,R}} (m z) \right].
\end{eqnarray}
The coefficients are fixed by the boundary conditions given in
Eqs. (\ref{eq:dmassbc}) and (\ref{eq:majmassbc}), and by imposing
canonical normalizations for the KK modes in the 4D effective
theory. The resulting mass spectrum for the lifted zero modes can
be modified by changing the $c_{L,R}$ parameters, and the
strengths of the boundary terms $M_D$, and $M$.

Accomplishing the quark splittings is slightly more complicated,
and involves adding induced kinetic terms on the boundaries, or
mixings with localized fermions.  However, the splittings of the
quarks can be achieved as well.

The final result is a spectrum which agrees with the standard
model.  It is possible to give the lighter fermions the right
masses while all of them are strongly localized on the UV brane.
This is ideal, since the distortion of the gauge boson wave
functions towards the IR brane due to the warp factor will shift
effective couplings to the fermion currents.  The third generation
of quarks poses a difficulty, however.  To get a large enough mass
for the top quark, the $(t,b)_L$ doublet must be localized towards
the IR brane.  This poses some difficulty for matching on to the
measured value of the $Z\bar{b} b$ coupling~\cite{Matt}.  This
will be discussed further in the next lecture.

\section{Electroweak Precision Observables For \\ General BSM and Extra Dimensional Theories}
\label{sec:EW} \setcounter{equation}{0}

This lecture dealing with electroweak~(EW) observables in extra
dimensional theories will be split into two main lines of
discussion. The first part of this lecture will be dealing with
how to calculate electroweak precision observables from the
effective Lagrangian point of view.  This part of the lecture will
be completely independent of extra dimensions and can be applied
to any weakly coupled model of beyond the SM~(BSM) physics.  After
this self contained introduction to EW observables in BSM physics
we will move on to analyzing where generic corrections to SM EW
observables come from in extra dimensional models.  The focus will
then be shifted to the particular cases of EW observables in RS
models and the Higgsless models.  The Higgsless models will be
discussed in detail at the end of this section showing how they
can both be ruled out from EW observables in their original
incarnations; as well as how these models can be made consistent
with EW fits when the original models are modified.

\subsection{EW Observables and the Effective Lagrangian Point of View}\label{ewsecprog}

There have already been a series of lectures on EW precision tests
by James Wells during this TASI~\cite{tasiwells}, however this
lecture will give a slightly different point of view that we find
very useful in practice.  We will follow closely here the analysis
done by Burgess et. al. in~\cite{Burgess:1993vc} but will attempt
to give a mostly self contained presentation.  There is a very
generic program that can be applied to electroweak precision
observables, and can be divided in the following way.

\begin{itemize}
\item First one needs to write down the most general effective
Lagrangian allowed by the symmetries that are unbroken at the
relevant energy scale (for instance for LEP I electroweak
precision observables this would be $M_Z$).

\item The second step after writing down the most general
Lagrangian is to fix the coefficients for this effective
Lagrangian for whatever model you are interested in analyzing.

\item Once this work has been done the third and final step is to
calculate the expressions for the desired observables in terms of
the input parameters and the effective Lagrangian coefficients.

\end{itemize}
The desired observables for the particular model in question,
$\mathcal{O}^{\mathrm{model}}$, are best usually to be expressed
as
$\mathcal{O}^{\mathrm{model}}=\mathcal{O}^{\mathrm{SM}}+\delta\mathcal{O}^{\mathrm{model}}$.
The reason for expressing the observables for a particular model
in this form is that the corrections
$\delta\mathcal{O}^{\mathrm{model}}$ are usually small (or you
better hope so if this is your model you are studying). One can
then include the SM loop contributions into
$\mathcal{O}^{\mathrm{SM}}$ and only deal with the tree-level
$\delta\mathcal{O}^{\mathrm{model}}$.  This is not always the case
as there are cases of BSM physics where the tree-level
contribution vanishes and loop level contributions must be
calculated, for instance as in~\cite{Hubisz:2005tx}, however it
can be successfully applied to a whole host of models.  At this
point once one has calculated the electroweak observables in the
model of interest one simply needs to perform a fit to the
experimental data and find the allowed regions of parameter space
for a particular model.

At this point several comments are in order after laying out such
a short and yet powerful program for testing new BSM physics
possibilities.  The first is that at no point in this program have
we made any reference to ``oblique" corrections or the  S, T, U
parameters~\cite{Peskin:1990zt} or the $\epsilon_1$, $\epsilon_2$,
$\epsilon_3$ parameters~\cite{Altarelli:1990zd}. Some students may
have heard of these various parameterizations and associated their
existence as the only calculations necessary for EW precision
tests (EWPT).  This is obviously not the case from the
prescription we have laid out so far, yet there will be a large
set of ``universal" corrections that can be included in oblique
corrections.  The choice of parametrization is a matter of
conventions and how you want to choose to express large sets of
common corrections.  For those who want more information on the
common parameterizations we refer the reader to some original
papers as well as useful
reviews~\cite{tasiwells,Peskin:1990zt,Altarelli:1990zd,Kennedy:1988rt,Kennedy:1988sn,Golden:1990ig,Holdom:1990tc,Matchev:2004yw}.

We will begin now with taking the Lagrangian at the energy scale
of the Z mass, $m_Z$, with the top quark and the Higgs integrated
out along with any other BSM particles as well.  The general form
of the gauge boson Lagrangian will be
\begin{equation}
\lag_{eff}=\lag_{SM}(\tilde{e}_i)+\hat{\lag}_{new},
\end{equation}
where $\lag_{SM}(\tilde{e}_i)$ is the ordinary SM Lagrangian with
SM loop effects taken into account.  The couplings of SM
Lagrangian $\tilde{e}_i$ do not take their usual numerical values,
because the new physics can contribute and has not been included.
For example $\frac{\tilde{e}^2}{4\pi}=\frac{1}{128}$ will not
necessarily hold due to the effects of $\lag_{new}$ which must be
taken into account.  The form of $\lag_{new}$ for the gauge boson
sector up to dimension $4$ is
\begin{eqnarray}\label{eqn:leff}
\hat{\lag}_{new}&=&-\frac{A}{4}\hat{F}_{\mu\nu}\hat{F}^{\mu\nu}-
\frac{B}{2}\hat{W}_{\mu\nu}\hat{W}^{\mu\nu}-\frac{C}{4}\hat{Z}_{\mu\nu}\hat{Z}^{\mu\nu}
+\frac{G}{2}\hat{F}_{\mu\nu}\hat{Z}^{\mu\nu}\nonumber\\ &&-w
\tilde{m_W}^2 \hat{W}_\mu^+ \hat{W}^{\mu\,-}-\frac{z}{2}
\tilde{m_Z}^2 \hat{Z}_\mu \hat{Z}^{\mu}.
\end{eqnarray}
The reason the gauge fields $\hat{A}$,$\hat{W}$ and $\hat{Z}$ are
hatted is because these are not canonically normalized fields.
There was no trick to the procedure of what to write down in
(\ref{eqn:leff}), we simply wrote down all dimension $\leq 4$
terms that are allowed by the remaining symmetries of the theory.
$\hat F_{\mu \nu}$ and $\hat Z_{\mu \nu}$ are the usual Abelian
field strengths, while the unbroken $U(1)_{EM}$ forces
\begin{equation}
\hat{W}_{\mu\nu}=D_\mu \hat{W}_\nu-D_\nu \hat{W}_\mu
\end{equation}
where
\begin{equation}
D_\mu \hat{W}_\nu=\partial_\mu \hat{W}_\nu+ i \tilde{e} A_\nu
\hat{W}_\mu.
\end{equation}
Let us assume for now that the rest of the SM Lagrangian in the
gauge sector is unchanged, that is
\begin{eqnarray}\label{restofsm}
\lag_{EM}&=& -\tilde{e} \sum_{i} \bar{f}_i \gamma^\mu Q_i f_i
\hat{A}_\mu,\nonumber\\
\lag_{CC}&=&-\frac{\tilde{e}}{\tilde{s}_W\sqrt{2}}\sum_{ij}\tilde{V}_{ij}
\bar{f}_i \gamma^\mu P_L f_j \hat{W}_\mu^+ +\mathrm{c.c},\nonumber\\
\lag_{NC}&=&-\frac{\tilde{e}}{\tilde{s}_W\tilde{c}_W} \sum_i
\bar{f}_i \gamma^\mu\left[T_{3i} P_L-Q_i \tilde{s}_W^2\right] f_i
\hat{Z}_\mu.
\end{eqnarray}
If (\ref{restofsm}) holds we have $6$ parameters $A,B,C,G,w$, and
$z$ however not all of them are physically observable.  The reason
for this is that field redefinitions of $W_\mu^a,$ $B_\mu$ and the
Higgs scalar $\phi$ can absorb $3$ of the six parameters.
Therefore in the end only $3$ combinations of these parameters
will appear in observables. Conventionally these three physical
combinations are called the familiar $S,T,$ and $U$ parameters,
which can be defined in terms of the parameters in
(\ref{eqn:leff}) as
\begin{eqnarray}\label{stuparam}
\alpha S &=& 4 s_W^2 c_W^2 (A-C- \frac{c_W^2-s_W^2}{c_W s_W}
G),\nonumber\\
\alpha T &=& w-z,\nonumber\\
\alpha U &=& 4 s_W^4
(A-\frac{B}{s_W^2}+\frac{c_W^2}{s_W^2}C-\frac{2c_W}{s_W}G)
\end{eqnarray}

Let us now begin to take into account the effect of the new
physics by rescaling the fields to get canonically normalized
gauge kinetic terms. Assuming the parameters $A,B,C,G,w,$ and $z$
are small we can go to canonically normalized kinetic terms with
the following rescaling
\begin{eqnarray}
\hat{A}_\mu &=& (1-\frac{A}{2}) A_\mu+ G Z_\mu \nonumber \\
\hat{W}_\mu &=& (1-\frac{B}{2}) W_\mu\nonumber \\
\hat{Z}_\mu &=& (1-\frac{C}{2}) Z_\mu.
\end{eqnarray}
This rescaling will bring the Lagrangian into the form
\begin{eqnarray}\label{eqn:effl2}
\lag_{eff}&=&-\frac{1}{4}F_{\mu\nu}F^{\mu\nu}-
\frac{1}{2}W_{\mu\nu}W^{\mu\nu}-\frac{1}{4}Z_{\mu\nu}Z^{\mu\nu}-(1+w-B)
\tilde{m_W}^2 W_\mu^+ W^{\mu\,-} \nonumber\\ &&-\frac{1}{2}(1+z-C)
\tilde{m_Z}^2 Z_\mu Z^{\mu} -\tilde{e} (1-\frac{A}{2}) \sum_{i}
\bar{f}_i \gamma^\mu Q_i f_i
A_\mu\nonumber\\
&&-\frac{\tilde{e}}{\tilde{s}_W\sqrt{2}}(1-\frac{B}{2})\sum_{ij}\tilde{V}_{ij}
\bar{f}_i \gamma^\mu P_L f_j W_\mu^+ +\mathrm{c.c}\nonumber\\
&&-\frac{\tilde{e}}{\tilde{s}_W\tilde{c}_W} (1-\frac{C}{2})\sum_i
\bar{f}_i \gamma^\mu\left[T_{3i} P_L-Q_i \tilde{s}_W^2+Q_i
\tilde{s}_W\tilde{c}_W G\right] f_i Z_\mu.
\end{eqnarray}
The Lagrangian (\ref{eqn:effl2}) depends on $3$ input parameters
$\tilde{e},\tilde{s}_W,$ and $\tilde{m}_Z$ (neglecting the CKM
elements). However we would like to trade these input parameters
for well measured quantities $\alpha, M_z, G_F$ that take on their
usual SM values. This will give us a relationship between the
tilded parameters $\tilde{e},\tilde{s}_W,\tilde{m}_Z,$ and the
measured observables $\alpha, M_z, G_F$, which can be further used
to express any other observables in terms of.

Starting with a general model as we have expressed so far, from
(\ref{eqn:effl2}) we have the relationship
\begin{equation}
4\pi\alpha = \tilde{e}^2 (1-A).
\end{equation}
In the SM this relationship takes the form $4\pi\alpha=e^2$, thus
equating the two will lead us to a relationship between
$\tilde{e}$ and $e$,
\begin{equation}
\tilde{e}=e\left(1+\frac{A}{2}\right).
\end{equation}
Similarly defining
\begin{equation}
M_Z^2=\tilde{m}_Z^2 (1+z-C)
\end{equation}
allows us to write the obvious relationship
\begin{equation}\label{tildemz}
\tilde{m}_z^2=M_z^2 (1-z+C).
\end{equation}

\begin{figure}[h]
\centerline{\includegraphics[width=0.4\hsize]{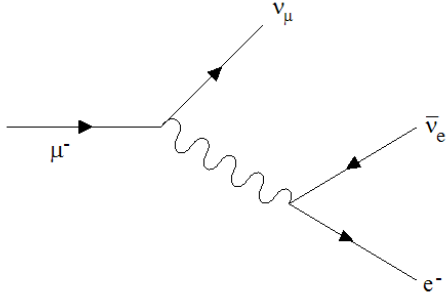}}
\caption{The process in the SM which contributes to
$G_F$.}\label{mudecay}
\end{figure}

We next turn our attention to $G_F$, Fermi's constant, which is
measured via $\mu$-decay.  The relevant diagram is shown in
Figure~\ref{mudecay}.  Integrating out the W boson generates an
effective four fermion operator whose dimensionful coupling
constant is given by $G_F$. In the SM the expression defining
$G_F$ is
\begin{equation}\label{smgf}
\frac{G_F}{\sqrt{2}}=\frac{e^2}{8s_W^2 c_W^2 m_Z^2}.
\end{equation}
There will be two effects that alter this relationship from the
introduction of the new operators BSM.  The first is the modified
charged current interactions in (\ref{eqn:effl2}) and the second
will be the shift in the W propagator which comes from a shift in
the pole mass which can be read off from (\ref{eqn:effl2}) as
well.

\noindent With these modifications from the effective Lagrangian
we have the relation
\begin{equation}
\frac{G_F}{\sqrt{2}} = \frac{\tilde{e}^2 (1-B)}{8 \tilde{s}_W^2
\tilde{c}_W^2 \tilde{m}_Z^2 (1+w-B)} \approx\frac{\tilde{e}^2
(1-w)}{8 \tilde{s}_W^2 \tilde{c}_W^2 \tilde{m}_Z^2}.
\end{equation}
>From this relationship and taking (\ref{smgf}) as the definition
of $s_W$ we can express $\tilde{s}_W^2$ as
\begin{equation}\label{stilde}
\tilde{s}_W^2=s_W^2\left[1+\frac{c_W^2}{c_W^2-s_W^2}(A-C-w+z)\right].
\end{equation}

We have now fixed part of the Lagrangian based on our chosen input
parameters and by construction
\begin{eqnarray}
\lag_{Z}&=&-\frac{1}{2}M_Z^2 Z_\mu Z^\mu\nonumber\\
\lag_{EM}&=& -e\sum_{i} \bar{f}_i \gamma^\mu Q_i f_i A_\mu.
\end{eqnarray}
However now we can express other observables in terms of our
chosen input observables and find the predictions for the effects
of the new physics.  For instance the term in the Lagrangian
governing the mass of the W was of the form
\begin{equation}\label{mwlag}
\lag_{W}=-(1+w-B)\tilde{m}_W^2 W_\mu^+ W^{\mu-} = -(1+w-B)
\tilde{c}_W^2 \tilde{m}_Z^2 W_\mu^+ W^{\mu-}
\end{equation}
where we have used the relationship $\tilde{m}_w=\tilde{m}_z
\tilde{c}_W$ since these refer to the tilded (SM only) parameters
which continue to take their natural relations.  We have
previously fixed the relationship of the tilded parameters to the
input observables in (\ref{stilde}) and (\ref{tildemz}) therefore
we can re-express (\ref{mwlag}) as
\begin{eqnarray}
\lag_{W}&=&-(1+w-B)(1-z+C)\left[1-\frac{s_W^2}{c_W^2-s_W^2}(A-C-w+z)\right]
M_Z^2 c_W^2 W_\mu^+ W^{\mu-}\nonumber\\
&=&-M_Z^2
c_W^2\left[1-B+C+w-z-\frac{s_W^2}{c_W^2-s_W^2}(A-C-w+z)\right]W_\mu^+
W^{\mu-}
\end{eqnarray}
If we now change to the S,T,U parametrization given in
(\ref{stuparam}) then the prediction for the physical mass of the
W as a function of the new physics is given by
\begin{equation}
(M_W^2)_{phys}=(M_W^2)_{SM}\left[1-\frac{\alpha S}{2
(c_W^2-s_W^2)}+\frac{c_W^2\alpha T}{c_W^2-s_W^2}+\frac{\alpha U}{4
s_W^2}\right].
\end{equation}

We can also analyze how other sectors of the Lagrangian are
changed by our choice of input parameters and what the effects are
for other observables.  In the charged current sector the
Lagrangian now takes the form
\begin{equation}
\lag_{CC}=-\frac{e}{\sqrt{2}s_W}\left(1+\frac{1}{2}\left[A-B-\frac{c_w^2}{c_W^2-s_W^2}\left(A-C-w+z\right)\right]\right)
\sum_{ij}V_{ij}\bar{f}_i \gamma^\mu P_L f_j W^{\mu+}+h.c.,
\end{equation}
which can be rewritten with the help of $S,T,$ and $U$ as
\begin{equation}\label{ccshift}
\lag_{CC}=-\frac{e}{\sqrt{2}s_W}\left(1-\frac{\alpha
S}{4(c_W^2-s_W^2)}+\frac{c_W^2 \alpha T}{2
(c_W^2-s_W^2)}+\frac{\alpha U}{8 s_W^2}\right)
\sum_{ij}V_{ij}\bar{f}_i \gamma^\mu P_L f_j W^{\mu+}+h.c.
\end{equation}
This now allows us to define a shifted coupling for charged
current interactions as $h_{ij}\equiv h_{ij}^{SM}+\delta h_{ij}$,
where $h_{ij}^{SM}=V_{ij}$ and the shift from new physics is
\begin{equation}
\delta h_{ij}=V_{ij}\left(-\frac{\alpha
S}{4(c_W^2-s_W^2)}+\frac{c_W^2 \alpha T}{2
(c_W^2-s_W^2)}+\frac{\alpha U}{8 s_W^2}\right)
\end{equation}
which can be read off from (\ref{ccshift}).  The neutral current
sector is also changed to the form
\begin{eqnarray}
\lag_{NC}=-\frac{e}{s_W c_W} (1+\frac{\alpha T}{2})\sum_i
\bar{f}_i \gamma^\mu\left[T_{3i} P_L-Q_i \left(s_W^2+\frac{\alpha
S}{4 (c_W^2-s_W^2)}-\frac{c_W^2 s_W^2 \alpha
T}{c_W^2-s_W^2}\right)\right] f_i Z_\mu.\nonumber\\
\end{eqnarray}
With this form of the neutral current we can define the couplings
$g_L\equiv g_L^{SM}+\delta g_L$ and $g_R\equiv g_R^{SM}+\delta
g_R$ analogously to the previously defined $h_{ij}$, where
\begin{eqnarray}\label{ncc}
g_L^{SM}&=& T_{3i}-Q_i s_W^2,\nonumber\\
g_R^{SM}&=& -Q_i s_W^2\nonumber,\\
&\mathrm{and}&\nonumber\\
 \delta g_{i L,R}&=&\frac{\alpha T}{2}
g_{i L,R}^{SM}-Q_i\left(\frac{\alpha
S}{4(c_W^2-s_W^2)}-\frac{c_W^2 s_W^2 \alpha
T}{c_W^2-s_W^2}\right).
\end{eqnarray}
With these definitions we can tackle other observables such as the
left right asymmetry at the $Z$-pole
\begin{equation}
A_{LR}=\frac{\Gamma(Z\rightarrow f_L
\bar{f}_L)-\Gamma(Z\rightarrow f_R \bar{f}_R)}{\Gamma(Z\rightarrow
f_L \bar{f}_L)+\Gamma(Z\rightarrow f_R \bar{f}_R)}.
\end{equation}
The left-right asymmetry can be rewritten in terms of the neutral
current couplings defined in (\ref{ncc}) as
\begin{eqnarray}
A_{LR}&=&\frac{g_{eL}^2-g_{eR}^2}{g_{eL}^2+g_{eR}^2}=A_{LR}^{SM}+
\frac{4g_{eL}^{SM}g_{eR}^{SM}}{(g_{eL}^{SM^2}+g_{eR}^{SM^2})^2}(g_{eR}^{SM}\delta
g_{eL}-g_{eL}^{SM}\delta g_{eR})\nonumber\\&=&A_{LR}^{SM}+
\frac{4g_{eL}^{SM}g_{eR}^{SM}}{(g_{eL}^{SM^2}+g_{eR}^{SM^2})^2}
(g_{eR}^{SM}-g_{eL}^{SM})\left(\frac{\alpha
S}{4(c_W^2-s_W^2)}-\frac{c_W^2 s_W^2\alpha T}{c_W^2-s_W^2}\right)
\end{eqnarray}

This program can be carried out for all other relevant observables
at the Z-pole and similarly one will get expressions that depend
only upon the oblique parameters $S,T,$ and $U$.  This conclusion
should be expected if we recall the form of the ``new" physics BSM
that we introduced in (\ref{eqn:leff}).  All corrections that we
assumed in (\ref{eqn:leff}) appeared in the gauge boson sector
only.  Sometimes this will be the case for a given model,
sometimes it won't.  The most common exceptions to new physics
appearing only obliquely are:
\begin{itemize}
\item exchange of heavy gauge bosons (KK modes!) gives an
additional contribution to $\mu$ decay as well as additional four
fermion operators\item mixing of heavy and light gauge bosons
could give a non-oblique contribution to the shift of $Zf\bar{f}$,
$Wf_i f_j$ couplings \item non universal fermion wave functions in
extra dimensions can also lead to the shift of $Zf\bar{f}$, $Wf_i
f_j$ couplings
\end{itemize}
while all of these have examples in extra dimensions there are
analogues in four dimensional models as well.  To account for
non-oblique corrections the program is no different in practice
than the one we have just discussed, the only difference being the
form of the effective Lagrangian (\ref{eqn:leff}).  Instead of
only including new operators in the gauge sector alone one could
include shifts in other sectors of the model up to whatever
dimension of operator was desired for accuracy.  This program has
been carried out in~\cite{Burgess:1993vc} up to and including
dimension five operators and we refer the interested reader there
for more details.  Of course including all operators up to
dimension five will not necessarily yield all the interesting
physics since new four fermion operators are dimension six.
However, this process can be straightforwardly extended to include
a more general effective Lagrangian up to dimension six.  The
interested reader may also find useful the extended (beyond S,T,
and U) ``universal" parametrization by Barbieri et.
al.~\cite{BPRS} which can account for new four fermion operators.

Before we move on to investigating the specific effects of extra
dimensions on EW precision observables we will further discuss
oblique corrections in the Peskin-Takeuchi $S,T,$ and $U$
formalism since it is the most widely used.  In the beginning of
the effective Lagrangian approach (\ref{eqn:leff}) we started off
with the parameters $A,B,C,G,w,$ and $z$ and then expressed these
six parameters in terms of three physically measurable
combinations $S,T,$ and $U$.  The Peskin-Takeuchi formalism calls
the parameters that we used in (\ref{eqn:leff}) by slightly
different names,
\begin{eqnarray}\label{pt}
\lag_{new}&=&\frac{\Pi'_{\gamma\gamma}(0)}{4}F_{\mu\nu}F^{\mu\nu}+
\frac{\Pi'_{WW}(0)}{2}W_{\mu\nu}W^{\mu\nu}+\frac{\Pi'_{ZZ}(0)}{4}Z_{\mu\nu}Z^{\mu\nu}
-\frac{\Pi'_{\gamma Z}(0)}{2}F_{\mu\nu}Z^{\mu\nu}\nonumber\\
&&+\Pi_{WW}(0) W_\mu^+ W^{\mu\,-}+\frac{\Pi_{ZZ}(0)}{2} Z_\mu
Z^{\mu}.
\end{eqnarray}

These are clearly the same parameters as what we used in
(\ref{eqn:leff}), their alternative names come from the reference
to the SM loop calculations in the Peskin-Takeuchi
paper\cite{Peskin:1990zt}. These parameters simply represent
shifts in the propagators of the gauge boson.  Using this language
and defining
\begin{eqnarray}
&&\Pi_{WW}(0)=g^2 \Pi_{11}(0)\;\;\;\;\;\;\;\;\;\;\;\,\Pi'_{WW}=g^2\Pi'_{11}(0)\nonumber\\
&&\Pi_{ZZ}(0)=(g^2+g^{'2})\Pi_{33}(0)\;\;\Pi'_{ZZ}(0)=(g^2+g^{'2})(\Pi'_{33}(0)-2
s_W^2 \Pi'_{3Q}(0)+s_W^4 \Pi'_{QQ}(0))\nonumber\\
&&\Pi'_{\gamma\gamma}(0)=e^2\Pi'_{QQ}(0)\;\;\;\;\;\;\;\;\;\;\;\;\,
\Pi'_{\gamma Z}(0)=g g' (\Pi'_{3Q}(0)-s_W^2 \Pi'_{QQ})
\end{eqnarray}
we can express the $S,T,$ and $U$ parameters in terms of these
$\Pi$ as
\begin{eqnarray}\label{stupt}
S&=& 16\pi (\Pi'_{33}(0)-\Pi'_{3Q}(0))\nonumber\\
T&=& \frac{4\pi}{s_W^2 c_W^2
M_Z^2}\left(\Pi_{11}(0)-\Pi_{33}(0)\right)\nonumber\\
U&=&16\pi(\Pi'_{11}(0)-\Pi'_{33}(0)).
\end{eqnarray}

We have now summarized a basic program that has been applied
previously to a wide range of models from little Higgs
models~\cite{Csaki:2002qg} to extra dimensional
ones~\cite{Csaki:2002gy,Csaki:2002bz} and we hope the reader can
apply it to any model of their own interest.

\subsection{Electroweak Precision and Extra
Dimensions}\label{ewsecxdim}

With the formalism set up in Section~\ref{ewsecprog} we may now
move forward with showing in detail how we implement this for
extra dimensional models.  For our first example let us assume
that the fermions will be localized at some given point. An
important point we should make is that matching between $5D$ and
$4D$ couplings is a convention.  However when choosing your
convention you want to make sure that your expressions are the
simplest such that it minimizes the amount of work for you to do.
For instance in a simple flat extra dimension the matching
condition between $4D$ and $5D$ gauge couplings usually is
\begin{equation}\label{matchg}
\frac{1}{g^2}=\frac{R}{g_5^2},\;
\mathrm{and}\;\frac{1}{g^{'2}}=\frac{R}{g_5^{'2}}.
\end{equation}
However if fermions are localized at one point, for instance
$y=0$, the coupling of fermions to gauge bosons is not necessarily
given by $g$ and $g'$ but rather the $4D$ effective Lagrangian is
of the form
\begin{equation}
\lag\supset -\int \frac{1}{4} F_{\mu\nu}F^{\mu\nu} dy+g_5
\bar{\psi}(0)\gamma_\mu \tilde{A}^\mu(0) \psi(0).
\end{equation}
If we canonically normalize the gauge field $A_\mu$ this in turn
shifts the interaction term to the form
\begin{equation}\label{intterm}
\lag_{int}=\frac{g_5}{\sqrt{R}} \bar{\psi}(0) \gamma_\mu \psi(0)
A^\mu(0),
\end{equation}
where the coupling will be given by $g A^\mu(0)$ if we match the
gauge coupling as in (\ref{matchg}).  To ensure that
\begin{bf}all\end{bf} corrections in this example will be oblique
requires that we need to pick the wave function at the location of
the fermions to be $1$ (if we use this simple matching
relationship).  We note here that this is not always possible in
the case that the fermions are localized at different points in
the extra dimension.  In this case the choice of the wave function
will then tell us how to completely calculate the $S,T,$ and $U$
parameters. The effective wave function for the light modes will
be
\begin{eqnarray}\label{effwf}
-\frac{1}{4}\int \vert\psi_\gamma (z)\vert^2 dz \,
F_{\mu\nu}F^{\mu\nu} -\frac{1}{2}\int \vert\psi_z (z)\vert^2 dz \,
Z_{\mu\nu}Z^{\mu\nu} -\frac{1}{2}\int \vert\psi_W (z)\vert^2 dz
\,W^{+}_{\mu\nu}W^{\mu\nu-}\nonumber\\
+\int dz \vert \partial_z \psi_W\vert^2 W^+_\mu W^{-\mu}+\int dz
\vert
\partial_z \psi_Z\vert^2 Z_\mu Z^\mu
+\mathrm{any}\,\mathrm{mass}\,\mathrm{from}
\,\mathrm{Higgs}\,\mathrm{terms}
\end{eqnarray}
Equation (\ref{effwf}) simply comes from the generic KK
decomposition into $4D$ fields. Beginning with (after setting
$A_5=0$ as previously discussed)
\begin{equation}
-\frac{1}{4}F_{MN}F^{MN}dz=-\frac{1}{4}\int
F_{\mu\nu}F^{\mu\nu}dz-\frac{1}{2}\int(\partial_5 A_\mu)^2 dz
\end{equation}
and writing
\begin{equation}
A_\mu(x,z)=A_\mu^0(x)\psi_A^0(z)+ \mathrm{KK\,modes}
\end{equation}
we get contributions to the Lagrangian of the form
\begin{equation}\label{kkcontrib}
-\frac{1}{4} \int dz \vert \psi_A(z)\vert^2 (\partial_\mu
A^0_\nu(x)-\partial_\nu A_\mu^0(x))^2+\frac{1}{2}\int dz \vert
\partial_z \psi_A(z)\vert^2 A^0_\mu(x) A^{0\mu}(x).
\end{equation}
As one can see from the last term in (\ref{kkcontrib}) a non-flat
wave function $\psi^0_A(z)$ will contribute to a $4D$ mass term!
With these, the expression for the  various $\Pi$'s will be
\begin{eqnarray}
g^2 \Pi_{11}(0)&=&\int \vert \partial_z \psi_W(z)\vert^2
dz\nonumber\\
(g^2+g'^2) \Pi_{33}(0)&=&\int \vert\partial_z \psi_Z(z)\vert^2\ dz \nonumber\\
1-g^2\Pi'_{11}(0)&=&\int \vert\psi_W(z)\vert^2 dz\nonumber\\
1-(g^2+g'^{2})\Pi'_{33}(0)&=&\int\vert\psi_Z(z)\vert^2 dz.
\end{eqnarray}
Now it is simply a matter of substituting these expressions into
(\ref{stupt}) to get the $S,T,$ and $U$ parameters.

For our next example we will look at the Randall-Sundrum model
with all the fermions as well as the Higgs on the TeV brane, but
with the gauge fields propagating in the bulk (this is
\begin{bf}not\end{bf} the most interesting case but probably the
simplest to actually calculate\cite{Csaki:2002gy}).  Starting with
the metric
\begin{equation}
ds^2=\frac{R}{z}(\eta_{\mu\nu} dx^\mu dx^\nu-dz^2)
\end{equation}
 the form of the action will be
\begin{eqnarray}
S_{5D}=\int
d^4x\int_{R}^{R'}dz\frac{R}{z}\left[-\frac{1}{2g_5^2}W_{MN}^+W^{MN-}-
\frac{1}{4}\left(\frac{1}{g_5^2}+\frac{1}{g_5^{'2}}\right)F_{MN}F^{MN}\right.\nonumber\\
\left.-\frac{1}{4(g_5^2+g_5^{'2})}Z_{MN}Z^{MN}+\frac{v^2}{4}\delta(z-R')\frac{R}{z}
W_M^+W^{M-}+\frac{v^2}{8}\delta(z-R')\frac{R}{z}Z_MZ^M\right].
\end{eqnarray}
Solving for the wave functions, the expressions for the masses and
wave functions of the $W$ and $Z$ are given by:
\begin{eqnarray}
M_W^2&=&\frac{1}{4}\frac{g_5^2}{R\log{R'/R}}\frac{R^2
v^2}{R^{'2}}\;\;\;\;\;\;\;\;\;
M_Z^2=\frac{1}{4}\frac{g_5^2+g_5^{'2}}{R\log{R'/R}}\frac{R^2
v^2}{R^{'2}}\nonumber\\
\psi_W^{(0)}&=&1+\frac{M_W^2}{4}\left[z^2-R^{'2}-2z^2\log{\frac{z}{R}}+2R^{'2}\log{\frac{R'}{R}}\right]\nonumber\\
\psi_Z^{(0)}&=&\psi_W^{(0)}(M_W\rightarrow M_Z).
\end{eqnarray}
The normalization of the wave functions is chosen such that
$\psi(R')=1$ (where the fermions are assumed to be localized).  We
note that the Higgs being on the TeV brane will not allow a flat
wave function for the $W$ and $Z$ in the bulk, which will give
rise to non-zero $S,T,$ and $U$ parameters.  Now we want to pick
our matching conditions so we can find the relevant $\Pi$'s. We
pick the simplest matching conditions
\begin{equation}
\frac{1}{g^2}=\frac{R\log{R'/R}}{g_5^2}\;\;\;\;\;\;\;\;\frac{1}{g^{'2}}=\frac{R\log{R'/R}}{g_5^{'2}}
\end{equation}
which leads to the following expressions
\begin{eqnarray}
\Pi'_{11}(0)=\Pi'_{33}(0)&=&-\frac{R^2}{R^{'2}}v^2\frac{1}{8}\left(2
R^{'2}\log{R'/R}-2R^{'2}+\frac{R^{'2}}{\log{R'/R}}+\dots\right),\\
\Pi_{11}(0)&=&-\frac{M_W^2}{2}\Pi'_{11}(0)\;\mathrm{and}\;\Pi_{33}(0)=-\frac{M_Z^2}{2}\Pi'_{33}(0).
\end{eqnarray}
With these $\Pi$'s we can easily calculate the $S,T,$ and $U$
parameters as we did before and we find that
\begin{equation}\label{sturs}
S=-4\pi f^2 R^{'2} \log{R'/R},\;\; T=-\frac{\pi}{2 c_W^2} f^2
R^{'2},\;\; U=0,
\end{equation}
where $f$ is the SM Higgs VEV.  We see from (\ref{sturs}) that the
contributions to $S$ and $T$ are quite sizeable.  However in this
case these are not the only corrections one must take into
account.  The KK modes of the gauge bosons localized on the TeV
brane will generate large corrections to $4$-fermi operators, in
particular $\mu$ decay and the definition of $G_F$.  We can
introduce a new parameter $V$ that measures the effect of the
$4$-fermi operators on the definition of $G_F$.  The effect of the
ordinary $W$ boson does not give the full $G_F$ but we will call
it $G_{F,W}$ which is given by
\begin{equation}\label{wcontrib}
4\sqrt{2} G_{F,W}\equiv\frac{1}{\frac{f^2}{4}+\Pi_{11}(0)}.
\end{equation}
The expression for the full $G_F$ defines the parameter $V$ as
\begin{equation}
G_F=G_{F,W} (1+V)
\end{equation}
where $V$ captures the effect of the additional KK modes of the
gauge bosons.  If we recall from Figure~\ref{mudecay}, the process
that defines $G_F$, we see that we need the full contribution to
the $W$ propagator.  This means that what we really need to do is
calculate the full $5D$ zero momentum brane to brane propagator
$\Delta_W (q=0,R',R')$.  We can calculate $V$ in terms of
$\Delta(0,R',R')$ and the W contribution (\ref{wcontrib}) which
gives
\begin{equation}
V=-\left(\Delta_W(0,R',R')+\frac{1}{\frac{f^2}{4}+\Pi_{11}(0)}\right)\left(
\frac{f^2}{4}+\Pi_{11}(0)\right).
\end{equation}
For the RS model the $W$ propagator will yield the result that
\begin{equation}\label{vrs}
V=\frac{g^2}{8}f^2 R^{'2}\log{\frac{R'}{R}}.
\end{equation}
With the $S,T,$ and $U$ results (\ref{sturs}) and the $V$
parameter (\ref{vrs}) we can now express all SM Z-pole observables
and find a bound on $R'$.  In this particular example one finds
that $\frac{1}{R'}>11\;\mathrm{TeV}$\cite{Csaki:2002gy}.  Of
course this is not the most interesting case since we would prefer
to have the fermions in the bulk (mostly localized to the Planck
brane). However even in this case there is still a strong
constraint on the model from the $T$ parameter.  The strong
constraint from the $T$ parameter in the original RS models with
the SM gauge symmetries are due to the absence of a custodial
$SU(2)$ symmetry. The solution to this is to put another $SU(2)_R$
gauge symmetry in the bulk of RS that is broken on the Planck
brane~\cite{ADMS}.  By the AdS/CFT correspondence this is
equivalent to adding a global custodial symmetry to the RS model
as discussed in Section~\ref{sec:lec2}.

\subsection{Electroweak Precision and Higgsless Models}

\begin{figure}[h]
\centerline{\includegraphics[angle=270,width=0.5\hsize]{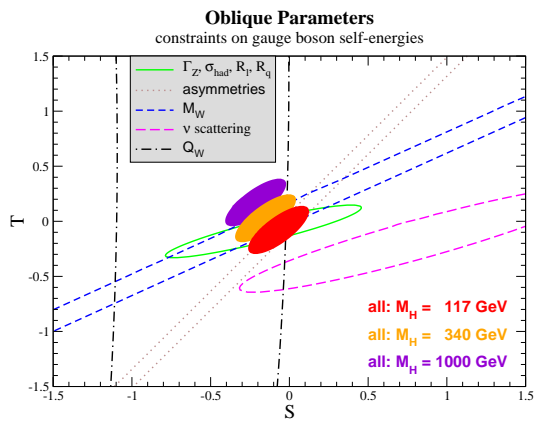}}
\caption[]{S-T Plot from PDG\cite{Eidelman:2004wy}}
\label{fig:stplot}
\end{figure}

We have seen from the previous Section~\ref{ewsecxdim} that to
leading order in $\log{R'/R}$ that $S=T=0$.  However the first
correction to the oblique parameters is
$\mathcal{O}(\frac{1}{\log{R'/R}}),$ which is relatively large and
one must check whether this is compatible with experimental
results.  We first analyze the Higgsless model discussed in
Section~\ref{sec:lec2} with the addition of Planck brane kinetic
terms
\begin{equation}
\frac{r}{4}W_{\mu\nu}^{L\,\,2}+\frac{r'}{4}\frac{1}{g_{5R}^2+\tilde{g}_5^2}(g_{5L}B_{\mu\nu}+\tilde{g}_5
W_{\mu\nu}^{R3})^2
\end{equation}
and TeV brane kinetic terms
\begin{equation}
-\frac{R'}{R}\left[\frac{\tau'}{4}B_{\mu\nu}^{2}+\frac{\tau}{4}\frac{1}{g_{5R}^2+g_{5L}^2}(g_{5L}W^L_{\mu\nu}+g_{5L}
W_{\mu\nu}^{R})^2\right].
\end{equation}
We perform the matching calculation and carry out the program that
we have been discussing and find the following approximate
expressions for $S$ and $T$ to leading order in $\tau$,
\begin{eqnarray}
S&\approx&\frac{6\pi}{g^2\log{\frac{R'}{R}}}\frac{2}{1+\frac{g_{5R}^2}{g_{5L}^2}}\frac{1}{1+\frac{r}{R\log{R'/R}}}
\left(1+\frac{4}{3}\frac{\tau}{R}\right)\nonumber\\
T&\approx&0.
\end{eqnarray}
However $M_W^2$ is given by
\begin{equation}
M_W^2\approx\frac{2}{1+\frac{g_{5R}^2}{g_{5L}^2}}\frac{1}{1+\frac{r}{R\log{R'/R}}}\frac{1}{R^{'2}\log{R'/R}}
\left(1-\frac{g_{5R}^2}{g_{5R}^2+g_{5L}^2}\frac{\tau}{r+R\log{R'/R}}\right).
\end{equation}
The experimental constraints from the oblique parameters are
\begin{eqnarray}
S=-0.13\pm0.10\nonumber\\
T=-0.17\pm0.11\nonumber\\
U=0.22\pm0.12
\end{eqnarray}
for a $117$ GeV Higgs with $1$ $\sigma$ error bars
given\cite{Eidelman:2004wy}. However the values are also
correlated in the usual S-T plot as shown in
Figure~\ref{fig:stplot}.
In the Higgsless models for $g_{5L}=g_{5R}$ and no induced kinetic
terms the results for $S$ and $T$ are
\begin{equation}
S=1.15\;\;\;\;\;\;T=0
\end{equation}
obviously completely excluding this form of the model\cite{CCGT}.

This isn't the end of the story since there is a
solution~\cite{CuringIlls} to the $S$ problem which has additional
beneficial side-effects. It has been known for a long time in
Randall-Sundrum (RS) models with a Higgs that the effective $S$
parameter is large and negative \cite{Csaki:2002gy} if the
fermions are localized on the TeV brane as originally proposed.
When the fermions are localized on the Planck brane the
contribution to $S$ is positive, and so for some intermediate
localization the $S$ parameter vanishes, as first pointed out for
RS models by Agashe et al.\cite{ADMS}. The reason for this is
fairly simple.  Since the $W$ and $Z$ wavefunctions are
approximately flat, and the gauge KK mode wavefunctions are
orthogonal to them, when the fermion wavefunctions are also
approximately flat the overlap of a gauge KK mode with two
fermions will approximately vanish. Since it is the coupling of
the gauge KK modes to the fermions that induces a shift in the $S$
parameter, for approximately flat fermion wavefunctions the $S$
parameter must be small. Note that not only does reducing the
coupling to gauge KK modes reduce the $S$ parameter, it also
weakens the experimental constraints on the existence of light KK
modes. This case of delocalized bulk fermions is not covered by
the no-go theorem of~\cite{BPRS}, since there it was assumed that
the fermions are localized on the Planck brane.

In order to quantify these statements, it is sufficient to
consider a toy model where all the three families of fermions are
massless and have a universal delocalized profile in the bulk.
Before showing some numerical results, it is useful to understand
the analytical behavior of $S$ in interesting limits. For fermions
almost localized on the Planck brane, it is possible to expand the
result for the $S$-parameter in powers of $(R/R')^{2c_L-1} \ll 1$.
The leading terms, also expanding in powers of $1/\log$, are:

\beq S = \frac{6 \pi}{g^2\, \log \frac{R'}{R}} \left( 1 -
\frac{4}{3} \frac{2c_L-1}{3-2c_L} \left(
\frac{R}{R'}\right)^{2c_L-1} \log \frac{R'}{R} \right)~, \eeq
\label{eq:Splanck} and $U \approx T \approx 0$. The above formula
is actually valid for $1/2 < c_L < 3/2$. For $c_L>3/2$ the
corrections are of order $(R'/R)^2$ and numerically negligible. As
we can see, as soon as the fermion wave function starts leaking
into the bulk, $S$ decreases.

Another interesting limit is when the profile is almost flat, $c_L
\approx 1/2$. In this case, the leading contributions to $S$ are:

\beq S =  \frac{2 \pi}{g^2\, \log \frac{R'}{R}} \left( 1 + (2 c_L
-1)\, \log \frac{R'}{R} + \mathcal{O} \left((2c_L-1)^2
\right)\right)~. \eeq In the flat limit $c_L=1/2$, $S$ is already
suppressed by a factor of 3 with respect to the Planck brane
localization case. Moreover, the leading terms cancel out for:
\beq c_L = \frac{1}{2} - \frac{1}{2\, \log \frac{R'}{R}} \approx
0.487~. \eeq

For $c_L<1/2$, $S$ becomes large and negative and, in the limit of
TeV brane localized fermions ($c_L \ll 1/2$):

\beq S =  - \frac{16 \pi}{g^2} \frac{1-2 c_L}{5-2 c_L}~, \eeq
while, in the limit $c_L\rightarrow - \infty$:
\begin{eqnarray}
T&\rightarrow& \frac{2 \pi}{g^2\, \log \frac{R'}{R}} (1 + \tan^2
\theta_W) \approx 0.5~,\\
U&\rightarrow& - \frac{8 \pi}{g^2\, \log \frac{R'}{R}}
\frac{\tan^2 \theta_W}{2 + \tan^2 \theta_W} \frac{1}{c_L} \approx
0~.
\end{eqnarray}

\begin{figure}[h]
\begin{center}
\includegraphics[width=0.5\hsize]{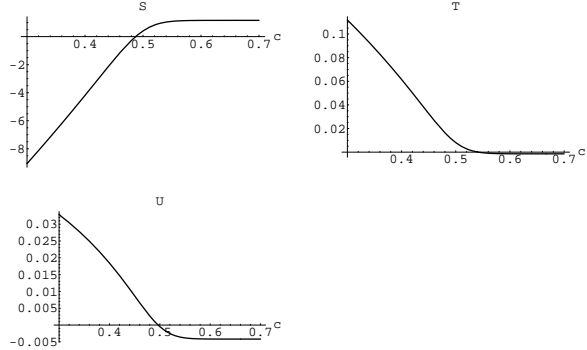}
\end{center}
\caption{ Plots of the oblique parameters as function of the bulk
mass of the reference fermion. The values on the right correspond
to localization on the Planck brane. $S$ vanishes for $c=0.487$. }
\label{fig:STUvsC}
\end{figure}

In Fig.~\ref{fig:STUvsC} we show the numerical results for the
oblique parameters as function of $c_L$. We can see that, after
vanishing for $c_L \approx 1/2$, $S$ becomes negative and large,
while $T$ and $U$ remain smaller. With $R$ chosen to be the
inverse Planck scale, the first KK resonance appears around
$1.2$~TeV, but for larger values of $R$ this scale can be safely
reduced down below a TeV.  Such resonances will be weakly coupled
to almost flat fermions and can easily avoid the strong bounds
from direct searches at LEP or Tevatron. If we are imagining that
the AdS space is a dual description of an approximate conformal
field theory (CFT), then $1/R$ is the scale where the CFT is no
longer approximately conformal and perhaps becomes asymptotically
free. Thus it is quite reasonable that the scale $1/R$ would be
much smaller than the Planck scale.

\begin{figure}[h]
\begin{center}
\includegraphics[width=0.7\hsize]{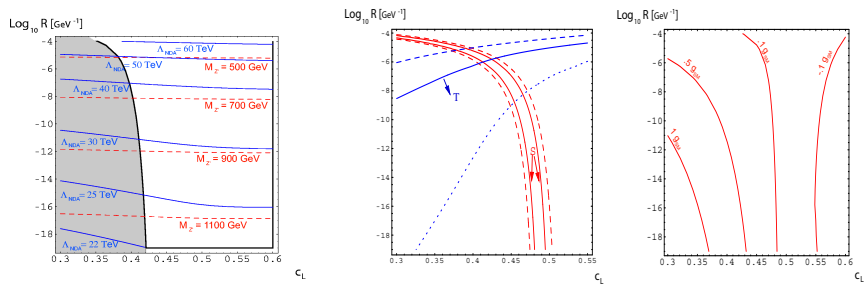}
\end{center}
\caption{In the left plot we show the contour plots of $\Lambda_{\rm NDA}$ (solid blue lines)
and $M_{Z^{(1)}}$ (dashed red lines) in the parameter space
$c_L$--$R$. The shaded region is excluded by direct searches of
light $Z^\prime$ at LEP. In the center, the
contours of $S$ (red), for $|S|=0.25$
(solid) and $0.5$ (dashed) and $T$ (blue), for  $|T|=0.1$
(dotted), $0.3$ (solid) and $0.5$ (dashed), as function of $c_L$
and $R$ are shown. On the right, contours for the generic suppression of
fermion couplings to the first resonance with respect to the SM
value can be seen. The region for $c_L$, allowed by
$S$, is between $0.43\pm 0.5$, where the couplings are suppressed
at least by a factor of 10.}
\label{fig:NDA}
\end{figure}

In Fig.~\ref{fig:NDA} we have plotted the value of the NDA scale
(\ref{NDA}) as well as the mass of the first resonance in the
$(c_L-R)$ plane. Increasing $R$ also affects the oblique
corrections. However, while it is always possible to reduce $S$ by
delocalizing the fermions, $T$ increases and puts a limit on how
far $R$ can be raised. One can also see from Fig.~\ref{fig:NDA}
that in the region where $|S|<0.25$, the coupling of the first
resonance with the light fermions is generically suppressed to
less than $10\%$ of the SM value. This means that the LEP bound of
$2$~TeV for SM--like $Z^\prime$ is also decreased by a factor of
10 at least (the correction to the differential cross section is
roughly proportional to $g^2/M_{Z'}^2$). In the end, values of $R$
as large as $10^{-7}$~GeV$^{-1}$ are allowed, where the resonance
masses are around $600$~GeV. So, even if, following the analysis
of~\cite{Papucci}, we take into account a factor of roughly $1/4$
in the NDA scale, we see that the appearance of strong coupling
regime can be delayed up to $10$~TeV.  At the LHC it will be very
difficult to probe $WW$ scattering above 3 TeV.

%

The major challenge facing Higgsless models is the incorporation
of the third family of quarks. There is a tension~\cite{BN,ADMS}
in obtaining a large top quark mass without deviating from the
observed bottom couplings with the $Z$. It can be seen in the
following way. The top quark mass is proportional both to the
Dirac mixing $M_D$ on the TeV brane and the overall scale of the
extra dimension set by $1/R'$. For $c_L\sim 0.5$ (or larger) it is
in fact impossible to obtain a heavy enough top quark mass (at
least for $g_{5R}=g_{5L}$). The reason is that for $M_DR' \gg 1$
the light mode mass saturates at
\begin{equation} m_{top}^2 \sim \frac{2}{R'^2 \log
\frac{R'}{R}}\,,
\end{equation}
which gives for this case $m_{top}\leq \sqrt{2} M_W$. Thus one
needs to localize the top and the bottom quarks closer to the TeV
brane. However, even in this case a sizable Dirac mass term on the
TeV brane is needed to obtain a heavy enough top quark. The
consequence of this mass term is the boundary condition for the
bottom quarks
\begin{equation}
\chi_{bR}= M_D R'\, \chi_{bL}.
\end{equation}
This implies that if $M_D R' \sim 1$ then the left handed bottom
quark has a sizable component also living in an $SU(2)_R$
multiplet, which however has a coupling to the $Z$ that is
different from the SM value. Thus there will be a large deviation
in the $Zb_L\bar{b}_L$. Note, that the same deviation will not
appear in the $Zb_R\bar{b}_R$ coupling, since the extra kinetic
term introduced on the Planck brane to split  top and bottom will
imply that the right handed $b$ lives mostly in the induced
fermion on the Planck brane which has the correct coupling to the
$Z$.

The only way of getting around this problem would be to raise the
value of $1/R'$, and thus lower the necessary mixing on the TeV
brane needed to obtain a heavy top quark. One way of raising the
value of $1/R'$ is by increasing the ratio $g_{5R}/g_{5L}$ (at the
price of also making the gauge KK modes heavier and thus the
theory more strongly coupled). Another possibility for rasing the
value of $1/R'$ is to separate the physics responsible for
electroweak symmetry breaking from that responsible for the
generation of the top mass. In technicolor models this is usually
achieved by introducing a new strong interaction called topcolor.
In the extra dimensional setup this would correspond to adding two
separate AdS$_5$ bulks, which meet at the Planck
brane~\cite{Matt}. One bulk would then be mostly responsible for
electroweak symmetry breaking, the other for generating the top
mass. The details of such models have been worked out
in~\cite{Matt}. The main consequences of such models would be the
necessary appearance of an isotriplet pseudo-Goldstone boson
called the top-pion, and depending on the detailed implementation
of the model there could also be a scalar particle (called the
top-Higgs) appearing. This top-Higgs would however not be playing
a major role in the unitarization of the gauge boson scattering
amplitudes, but rather serve as the source for the top-mass only.

\section{Conclusions}

We have attempted to give an introduction to the uses of extra
dimensions for electroweak physics. Clearly there are many other
very interesting models in this field besides the higgsless
theories, which we were not able to cover in these lectures. Here
is a partial list of the most prominent models of new electroweak
phenomenology from extra dimensions not covered here:
\begin{itemize}
\item The Randall-Sundrum model
with custodial SU(2)~\cite{ADMS}, which can be thought of as the simplest
implementation of the composite Higgs idea of Georgi and Kaplan.

\item The hierarchy problem could also be possibly solved if the
Higgs was secretly an extra dimensional component ($A_5$) of the gauge
field. These theories are usually referred to as
models with ``gauge-higgs unification''~\cite{gaugehiggs}
and were also the original inspirations for the little Higgs models
of~\cite{lh}.

\item Most recently~\cite{ACP}, the above two approaches have been
combined to get a model with gauge-higgs unification in warped
space, with the $A_5$ component localized on the TeV brane. This
way the Higgs would be a composite pseudo-Goldstone boson,
explaining its lightness compared to the TeV scale. To date these
are probably the most successful models of electroweak
phenomenology from extra dimensions, which may even incorporate a
successful unification of the gauge couplings~\cite{ACS}.
\end{itemize}

We can only guess that there must be many other interesting models
of the TeV scale that no one has thought of yet. We are all hoping
that the guessing will end abruptly about 2-3 years from now, and
at TASI 2009 we will already be lecturing about the definitive
theory of electroweak symmetry breaking (and start a new guessing
game of what lies beyond)...

\section*{Acknowledgments}
We thank K.T. Mahanthappa, John Terning, Carlos Wagner and Dieter
Zeppenfeld for organizing a stimulating TASI. C.C. thanks Giacomo
Cacciapaglia, Christophe Grojean, Hitoshi Murayama, Luigi Pilo,
Matt Reece, Yuri Shirman and John Terning for collaborations which
form much of the basis of these lectures. We also thank Giacomo
Cacciapaglia, Cristophe Grojean, and Matt Reece for comments on
the manuscript. This research is supported in part by the DOE OJI
grant DE-FG02-01ER41206 and in part by the NSF grants PHY-0139738
and PHY-0098631.

\end{document}